\documentclass{article}
\usepackage{amsfonts}
\usepackage{amssymb}
\usepackage{graphicx}
\usepackage{amsmath}

\input{tcilatex}

\begin{document}

\title{Clifford Tetrads, Null Zig Zags, and Quantum Gravity\thanks{%
This paper gives the derivations of the results I reported at the PIMS
conference entitled \textquotedblleft Brane World and Supersymmetry" in
July, 2002 at Vancouver, B.C. It also contains new results on spin-gravity
coupling, on how a topologically-nontrivial distribution of vacuum spinors
removes singularities and divergences, and how the amplitude of the vacuum
spinors determines the gravitational constant and the rate of cosmic
expansion.}}
\author{Marcus S. Cohen \\
Department of Mathematical Sciences\\
New Mexico State University\\
Las Cruces, New Mexico\\
marcus@nmsu.edu}
\date{\today }
\maketitle

\begin{abstract}
Quantum gravity has been so elusive because we have tried to approach it by
two paths which can never meet: standard quantum field theory and general
relativity. These contradict each other, not only in superdense regimes, but
also in the \emph{vacuum}, where the divergent zero-point energy would roll
up space to a point. The solution is to build in a regular, but
topologically nontrivial distribution of \emph{vacuum spinor fields} right
from the start. This opens up a straight road to quantum gravity, which we
map out here.

The gateway is covariance under the \emph{complexified Clifford algebra} of
our space-time manifold $\mathbb{M}$, and its spinor representations, which
Sachs dubbed the \emph{Einstein group,} $E$. The 16 generators of $E$
transformations obey both the \emph{Lie algebra} of\emph{\ Spin}$^{c}$\emph{%
-4}, and the Clifford (SUSY) algebra of $\mathbb{M}$. We derive Einstein's
field equations from the simplest $E$-invariant Lagrangian density, $%
\mathcal{L}_{g}$. $\mathcal{L}_{g}$ contains effective \emph{electroweak}
and \emph{gravitostrong field} actions, as well as Dirac actions for the
matter spinors. On microscales the massive Dirac propagator resolves into a
sum over \emph{null zig-zags}. On macroscales, we see the \emph{%
energy-momentum current,} $\ast T$, and the resulting Einstein curvature, $G$%
.

For massive particles, $\ast T$ flows in the \textquotedblleft cosmic
time\textquotedblright\ direction---\emph{centri-fugally} in an expanding
universe. Neighboring centrifugal currents of $\ast T$ present \emph{opposite%
} radiotemporal vorticities $G_{or}$ to the boundaries of each others'
worldtubes, so they \emph{advect,} i.e. \emph{attract,} as we show here by
integrating $\mathcal{L}_{g}$ by parts in the \emph{spinfluid} regime. This
boundary integral not only explains \emph{why} stress-energy is the source
for gravitational curvature, but also gives a \emph{value} for the
gravitational constant, $\kappa \left( T\right) $, that couples them. $%
\kappa $ turns out to depend on the dilation factor $T=y^{0}$, which enters
kinematically as \textquotedblleft imaginary time": the logradius of our
expanding Friedmann $3$-brane.

On the microscopic scale, \emph{quantum gravity} appears as the \emph{%
statistical mechanics} of the \emph{null zig-zag} rays of spinor fields in
imaginary time $T$. Our unified field/particle action $\mathcal{L}_{g}$ also
contains new couplings of gravitomagnetic fields to strong fields and weak
potentials. These \emph{predict new physical phenomena: Axial jets} of
nuclear decay products emitted with \emph{left helicity} along the axis of a
massive, spinning body.
\end{abstract}

\section{Overview: Vacuum Energy and Inertial Mass}

Quantum Field Theory (QFT) is \emph{incompatible} with General Relativity
(GR), not only at small scales or high densities, but in the \emph{vacuum!}
The problem is that the divergent QFT vacuum energy would produce enough
spacetime curvature to roll up our space to a point. The \emph{solution} is
a fundamental theory from which both QFT and GR derive, in different
regimes. We show here how quantum gravity emerges naturally from such a
theory: the Nonlinear Multispinor (NM) model \cite{capp}, \cite{spingeo}.

In this model, the particles emerge \cite{tqac}, \cite{strat} through \emph{%
dynamical symmetry breaking} of a topologically nontrivial \textquotedblleft
vacuum" solution. Their interactions are mediated by the phase perturbations
they leave in the residual vacuum around them.

We may summarize the role of the vacuum spinors in producing inertial and
gravitational mass \cite{spingeo} like this:

\begin{quote}
\emph{Nonlinear interaction of opposite-chirality components through the
vacuum spinor fields creates the inertial mass of a bispinor particle. It
also produces the gravitational interaction between particles.}
\end{quote}

Inertial mass from global interaction---and its corollary, gravitation, are 
\emph{results} we \emph{derive} from the NM model. We thus supply a \emph{%
mechanism} that embodies \emph{Mach's principle} (inertia from interaction
with the \textquotedblleft distant masses"), and Einstein's principle
(inertial mass equals gravitational mass). Let us start with the
\textquotedblleft big picture", and then zoom in to the microscopic level.

The new \textquotedblleft precision cosmology" \cite{smoot} has suddenly
given us a clear view of the universe in which we live. We live on a $3$%
\emph{-brane} (hypersurface) $S_{3}\left( a\right) $ whose (local) radius $%
a\left( x\right) $ is expanding with intrinsic clocktime: Minkowsky time, $%
t\equiv x^{0}$; the arclength travelled by a photon, projected to $\mathbb{S}%
_{3}\left( a_{\#}\right) $ \cite{im}. We take $S_{3}$ to be a \emph{closed}
hypersurface---a deformed, nonuniformly expanding 3-sphere.

If we could rise above our expanding spatial hypersurface $S_{3}$ and look
down on it from the direction $T=y^{0}$ of cosmic expansion, we would see a
patchwork quilt of matter concentrations connected by the fabric we call
\textquotedblleft space". It is becoming clear that space is not empty. It
is filled with \textquotedblleft dark energy", which includes the quantum
mechanical \emph{vacuum energy} and comprises over 70\% of the energy in the
universe \cite{sims}. On the macroscopic scale, this is what produces the
cosmological constant term that brings the average energy (almost exactly)
to the critical value needed to just close the universe.

Of what is this vacuum energy made? Before we get a microscopic view of the
vacuum, let us first zoom in on some of the matter concentrations that it
separates---starting with a single electron. The worldtube, $B_{4}$, of an
electron (or \emph{any} massive particle at rest) runs in the \emph{cosmic
time} direction, $T$, orthogonal to the $3$ spatial directions that span our 
\emph{spatial }$3$\emph{-brane,} $S_{3}\left( T\right) $.

The chiral components $\xi _{-}$ and $\eta _{-}$ of $e_{-}$ are \emph{%
spinors:} lightlike waves of internal $u\left( 1\right) \oplus su\left(
2\right) $ phase with \emph{definite helicity} (spin in the direction of
propagation). Since $\xi _{-}$ and $\eta _{-}$ have \emph{opposite helicities%
} but the \emph{same spins,} they must be \emph{%
counterpropagating---radially outward and inward for an electron at rest.}
Since the rest frame of a massive particle drifts \emph{slowly} outward with
cosmic expansion, there must be \emph{almost} as many inward
(\textquotedblleft backward" in $T$) as outward (\textquotedblleft forward")
lightlike ray segments within its worldtube \cite{ord}.

At the worldtube boundary $\partial B_{4}$, zigs are scattered into zags and
vice versa by nonlinear interactions with the vacuum spinors \cite{spingeo}.
It is these \emph{mass scatterings} \cite{penrose} that keep the lightlike
spinors of a massive bispinor particle confined to a timelike worldtube.
Energy-momentum, angular momentum, and all internal quantum numbers are
conserved at each mass scattering, where $4$ Lie-algebra phases of incoming
and outgoing spinors combine to give a \emph{scalar} contribution to the
action.

Mass scatterings, or \emph{Spin}$^{c}\emph{-4}$ \emph{resonances,} are the
multispinor analog of the \emph{Bragg resonances} (4-wave mixing) \cite%
{bragg} which produce the self-trapping that leads to soliton formation in
nonlinear optics. The same nonlinearity gives the \emph{soliton interactions}%
. These are most easily calculated for conservative fields by integrating
the Lagrangian density by parts over the boundary of the worldtube of an
accelerated particle. Matching the \emph{inner form} of the boundary
integral in the localized soliton fields to the \emph{outer form} in the
vacuum fields, as perturbed by source distributions, gives the curvature of
the particle's worldtube \cite{neu}.

We derive Einstein's field equations here by integrating a topological
Lagrangian by parts. The crucial step is to recognize that Einstein
curvature $G$ and energy-momentum $\ast T$ are different expressions for the 
\emph{same} flux, \emph{the spinfluid current\ }$3$\emph{\ form}. Inside the
worldtube, $B_{4}$, of a particle, this takes the form of the
stress-energy-momentum tensor \cite{landl}, $\ast T$. This is Noether
current under displacement of the \emph{boundary} $\partial B^{4}$ of the
worldtube. In the outer region, the spinfluid current takes the form of the 
\emph{Einstein curvature }\cite{mtw}, $G$. By matching the inner and outer
forms on the \emph{moving boundary} $\partial B_{4}\left( t\right) $, we
obtain Einstein's field equations.

We obtain the quantum mechanical form of this boundary integral by focusing
on a patch of boundary at a microscopic scale, where the matter current is
resolved into a sum over null zig-zags \cite{penrose}; mass-scatterings with
the vacuum fields, as perturbed by source distributions. Analyticity
conditions convert the \emph{statistical mechanics} of mass scatterings in
imaginary time or \emph{logradius} $T\equiv y^{0}$ to a quantum theory of
the mutual attraction of matter currents: \emph{quantum gravity}.

\section{Spinfluid Flow: The Dilation-Boost Current}

Conserved currents spring from \emph{invariances} under the group of spin
isometries of our spacetime manifold $\mathbb{M}$: the \emph{Einstein} ($E$) 
\emph{group} \cite{sachs}. Passive Einstein ($E_{P}$) transformations relate
the same physical state $\psi $ in different frames of reference. Active
Einstein ($E_{A}$) transformations change $\psi $ in a way that can be
undone by \emph{local,} \emph{path dependent} $E_{P}$ transformations. \emph{%
Spin curvatures,} or \emph{fields,} are the obstruction to \textquotedblleft
combing" $\psi $ to covariant constancy by \emph{any} (path-independent) $%
E_{P}$ transformation.

Cosmic expansion and boost-covariance demand that $E$ be \emph{nonunitary}%
---i.e. have \emph{Hermitian }$\left( H\right) $ as well as \emph{%
anti-Hermitian }$\left( aH\right) $ generators. The \emph{energy-momentum}
current, $\ast T$, is the \emph{Noether current of the spinor fields under
spacetime translations }\cite{landl}.\emph{\ }In an expanding universe \cite%
{im}, the net\emph{\ dilation flow} is\emph{\ centrifugally outward,} in the
direction $T$ of cosmic expansion. Cosmic time $T\equiv y^{0}$ enters
kinematically \cite{im}, \cite{penrose} as the \emph{imaginary} part of a 
\emph{complex time} variable $z^{0}\equiv x^{0}+iy^{0}$, where $x^{0}\equiv
t $ is \emph{Minkowsky time}. $y^{0}$ transforms like an \emph{energy. }The
imaginary parts $y^{j}$ of $z^{j}\equiv x^{j}+iy^{j}$ transform like $3$%
-momenta. The \emph{dilation current} is the \emph{energy density}, the
boost current is the momentum density. The $4$ complex variables $z^{\alpha
} $ are coordinates on the \emph{position-momentum phase space} \cite%
{spingeo}, \cite{im}:%
\begin{equation*}
\mathbb{M}\subset T^{\ast }\mathbb{M}\subset \mathbb{C}_{4}\text{;}
\end{equation*}%
the base space for the bundle of \emph{vacuum} $\oplus $ \emph{matter
spinors.} In the \emph{spinfluid regime}, the dilation-boost flow $y^{\alpha
}\left( x\right) $ is a function of Minkowsky-space position $x\equiv \left(
x^{0},x^{1},x^{2},x^{3}\right) \in \mathbb{M}_{\#}$.

The spin isometry group of $T^{\ast }\mathbb{M}$---the globalization of the
Poincar\'{e} group---is the \emph{Einstein group} \cite{sachs}%
\begin{equation*}
E\sim Spin^{c}\text{-4}\sim Gl\left( 2,\mathbb{C}\right) _{L}\times Gl\left(
2,\mathbb{C}\right) _{R}\text{,}
\end{equation*}%
the complexification of $\left( Spin\;4\right) \times U\left( 1\right) $---a
16-parameter Lie group. $E$ covers the \emph{conformal group}\footnote{%
The chiral $GL\left( 2,\mathbb{C}\right) $ presentation turns out to be
better suited than the twistor presentation for unifying Dirac mass with
gravitation---a fact first pointed out to me by Jaime Keller \cite{keller}.}
with one extra $U\left( 1\right) $---or \emph{electromagnetic} parameter.
The parameters $x^{\alpha }\in \left[ 0,4\pi a_{\#}\right] $ that multiply
the $aH$ generators are canonical $u\left( 1\right) \times su\left( 2,%
\mathbb{C}\right) $ \emph{translation} parameters on $\mathbb{M}_{\#}\equiv 
\mathbb{S}_{1}\times \mathbb{S}_{3}\left( a_{\#}\right) $, Penrose's \cite%
{penrose} conformal compactification of Minkowsky space. $\mathbb{S}%
_{3}\left( a_{\#}\right) $ is a reference $3$-sphere of radius $a_{\#}$. The
fundamental length unit $a_{\#}$ turns out \cite{im}, \cite{fried} to be the 
\emph{equilibrium radius} of the Friedmann solution \cite{im} (see Appendix).

We treat the \emph{vacuum} $\widehat{\mathbb{M}}\equiv \mathbb{M}\backslash
\cup D_{J}$ outside the worldtubes of massive particles as a \emph{spinfluid:%
} an inhomogeneous (but regular) distribution of $4$ path-dependent spinors $%
\mathbf{\psi }_{I}\left( x\right) $ and $4$ cospinors $\mathbf{\psi }%
^{I}\left( x\right) $, governed by a Lagrangian density, $\mathcal{L}_{g}$. $%
\mathcal{L}_{g}$ is invariant under the Lie group of passive ($E_{P}$)
changes in the background spin and spacetime frames, augmented by the
discrete involutions $P$ (space-reversal), $T$ (time reversal), and $C$
(charge reversal). A $PTC$ invariant inner product $\psi _{\mp }^{I}\psi
_{I}^{\pm }$ is made between spinors $\psi _{I}^{\pm }$ and \emph{cospinors} 
$\psi _{\pm }^{I}$ of \emph{Opposite Charge} ($u\left( 1\right) $ phase
shift with $T$), \emph{Parity} ($su\left( 2\right) $ phase advance along
rays), and \emph{Temorality} (inward, i.e. forward, or outward, i.e.
backward, propagation in $T$).

A good way to describe a physical state is by the \emph{active local} ($%
E_{A} $) transformation, which creates this state from the vacuum. Suppose
that each spinor field may be created from the \emph{vacuum} distribution $%
\widehat{\mathbf{\psi }}_{\pm }$ by (a path-dependent) active-local Einstein
($E_{A}$) transformation \cite{im}, \cite{xueg}. These act on the \emph{%
basis spinors} to create the \emph{moving spin frames} $g_{I}\left( x\right) 
$ written as $gl\left( 2,\mathbb{C}\right) $ matrices. Each spinor $\psi
_{I}^{\pm }$ and cospinor $\psi _{\mp }^{I}$ is expressed as a linear
combination of the two (\textquotedblleft spin-up" and \textquotedblleft
spin-down") basis spinors in its moving spin frame, with coefficients given
by the column or row \emph{spin vectors} $\mathbf{\psi }_{I}^{\pm }\left(
x\right) $ or $\mathbf{\psi }_{\pm }^{I}\left( x\right) $: 
\begin{equation}
\begin{array}{c}
\psi _{I}^{\pm }\left( x\right) =\exp \left( \frac{i}{2}\zeta _{I}^{\alpha
\pm }\left( x\right) \sigma _{\alpha }\right) \mathbf{\psi }_{I}^{\pm
}\equiv g_{I}^{\pm }\left( x\right) \mathbf{\psi }_{I}^{\pm }\text{,} \\ 
\psi _{\pm }^{I}\left( x\right) =\mathbf{\psi }_{\pm }^{I}\exp \left( \frac{i%
}{2}\zeta ^{I\alpha \pm }\left( x\right) \overline{\sigma }_{\alpha }\right)
\equiv \mathbf{\psi }_{\pm }^{I}g_{\pm }^{I}\left( x\right) \text{;} \\ 
\alpha =\left( 0,1,2,3\right) \text{,}%
\end{array}
\label{1}
\end{equation}%
where $\overline{\sigma }_{\alpha }\sim \left( \sigma _{0},-\sigma
_{1},-\sigma _{2},-\sigma _{3}\right) $ is the Lie algebra dual to $\sigma
_{\alpha }$ under the \emph{Clifford product}%
\begin{equation*}
\sigma _{\alpha }\overline{\sigma }_{\beta }+\sigma _{\beta }\overline{%
\sigma }_{\alpha }=2\eta _{\alpha \beta }
\end{equation*}%
for Minkowsky space. The $\pm $ signs indicate the \emph{charge} ($u\left(
1\right) $ phase shift) of the field. (We shall sometimes drop the charge
scripts below.) Spinors and cospinors must be \emph{varied independently} in
the Lagrangian, which must contain \emph{both} to be a \emph{scalar} under $%
E_{P}$.

In the \emph{geometrical-optics} (g.o.) regime, each spinor has a \emph{%
complex}, nonsingular (but perhaps path-dependent) phase (\ref{1}) with
(perhaps inexact) $gl\left( 2,\mathbb{C}\right) $ phase differential $%
\mathbf{d}\zeta _{I}^{\alpha }\left( x\right) $. In g.o. solutions, the
cospinor turns out to be \cite{spingeo} the $PTC$-reversed version of the
spinor, with the \emph{opposite} $gl\left( 2,\mathbb{C}\right) $ phase shift.

When you differentiate a spinor, you must also differentiate its moving spin
frame:%
\begin{equation}
\begin{array}{c}
\mathbf{d}\psi _{I}\left( x\right) =\mathbf{d}\left( g_{I}\left( x\right) 
\mathbf{\psi }_{I}\left( x\right) \right) =g_{I}\mathbf{d\psi }_{I}+\mathbf{d%
}g_{I}\mathbf{\psi }_{I} \\ 
\equiv g_{I}\left[ \mathbf{d}+g_{I}^{-1}\mathbf{d}g_{I}\right] \mathbf{\psi }%
_{I}\equiv g_{I}\left[ \mathbf{d}+\Omega _{I}\left( x\right) \right] \mathbf{%
\psi }_{I}\left( x\right) \equiv g_{I}\mathbf{\nabla \psi }_{I}\text{.}%
\end{array}
\label{2}
\end{equation}%
The \emph{spin connections} (vector potentials) 
\begin{equation}
g_{I}^{-1}\mathbf{d}g_{I}=\frac{i}{2}\mathbf{d}\zeta _{I}^{\alpha }\left(
x\right) \sigma _{\alpha }=\frac{i}{2}\left[ \mathbf{d}\theta _{I}^{\alpha
}\left( x\right) +i\mathbf{d}\varphi _{I}^{\alpha }\right] \sigma _{\alpha
}\equiv \Omega _{I}\left( x\right)  \label{3}
\end{equation}%
thus enter as $gl\left( 2,\mathbb{C}\right) $-valued $1$ forms into the 
\emph{covariant derivatives} of each spinor field: 
\begin{equation*}
\nabla _{\beta }\mathbf{\psi }_{I}\equiv \left( \partial _{\beta }+\Omega
_{I\beta }\left( x\right) \right) \mathbf{\psi }_{I}\text{.}
\end{equation*}%
The $\Omega _{I}\left( x\right) $ record the (perhaps path-dependent) phase
shift of each moving spin frame $g_{I}\left( x\right) $, in any direction at
point $x\in \mathbb{M}$ due to local sources---and of the global (vacuum)
distribution.

The simplest way to guaranty covariance of the wave equations for spinor, or 
\emph{spin-tensor} fields is to write the spacetime position vector $q\left(
x\right) $ from the origin, $x=0$, to the spacetime position point $q$, as
the $\emph{positio}$\emph{n quaternion}%
\begin{equation}
\begin{array}{c}
q\left( x\right) \equiv a\sigma _{0}\exp \left[ \frac{i}{a_{\#}}x^{\alpha
}\sigma _{\alpha }\right] \equiv a^{0}\left( x\right) \sigma
_{0}+ia^{j}\left( x\right) \sigma _{j}\in \mathbb{M}_{\#}\text{:} \\ 
\left( a^{0}\right) ^{2}+\left( a^{j}\right) ^{2}=a\qquad \text{(sum on }%
j=1,2,3\text{).}%
\end{array}
\label{4}
\end{equation}%
The point has an \emph{embedded radius}%
\begin{equation}
\begin{array}{c}
a\left( x\right) \equiv \exp \left[ \frac{1}{a_{\#}}y^{0}\left( x\right) %
\right] a_{\#}\equiv \gamma \left( x\right) a_{\#}\text{;} \\ 
\gamma \left( x\right) \equiv \frac{a\left( x\right) }{a_{\#}}%
\end{array}
\label{5}
\end{equation}%
is the (local) \emph{scale factor.}

For the stationery, homogeneous vacuum $\mathbb{M}_{\#}\equiv \mathbb{S}%
_{1}\times \mathbb{S}_{3}\left( a_{\#}\right) $, the \emph{left-invariant
spin connections} are the left-invariant Maurer-Cartan $1$ forms that derive
from \emph{right} action on $q\left( x\right) $ by the four canonical maps
of $\mathbb{M}_{\#}$ onto $U\left( 1\right) \times SU\left( 2\right) $ (and 
\emph{vice versa}):%
\begin{equation}
g^{\pm }\left( x\right) \equiv \exp \frac{i}{2a_{\#}}x^{\alpha }\sigma
_{\alpha }^{\pm }\text{;\qquad }\overline{g}^{\pm }\left( x\right) \equiv
\exp \frac{i}{2a_{\#}}x^{\alpha }\overline{\sigma }_{\alpha }^{\pm }\text{;}
\label{6}
\end{equation}%
where $\sigma _{\alpha }\equiv \left( \pm \sigma _{0},\sigma _{1},\sigma
_{2},\sigma _{3}\right) $ and $\bar{\sigma}_{\alpha }^{\pm }=\left( \pm
\sigma _{0},-\sigma _{1},-\sigma _{2},-\sigma _{3}\right) $.

On our \emph{Friedmann vacuum} $\widehat{\mathbb{M}}$, with time-dependent
scale factor $\frac{y^{0}\left( t\right) }{a_{\#}}=\gamma \left( t\right) $,
the left-invariant spin connections are%
\begin{equation}
\begin{array}{c}
\hat{\Omega}_{L}^{\pm }\equiv g_{\mp }^{L}\mathbf{d}g_{L}^{\pm }=\frac{i}{%
2a_{\#}}\sigma _{\alpha }^{\pm }e^{\alpha }-\frac{1}{2}\overset{\cdot }{%
\gamma }\sigma _{0}e^{0} \\ 
\hat{\Omega}_{R}^{\pm }\equiv g_{\mp }^{R}\mathbf{d}g_{R}^{\pm }=\frac{i}{%
2a_{\#}}\bar{\sigma}_{\alpha }^{\pm }e^{\alpha }-\frac{1}{2}\overset{\cdot }{%
\gamma }\sigma _{0}e^{0}\text{,}%
\end{array}
\label{7}
\end{equation}%
where%
\begin{equation*}
\begin{array}{c}
g_{L}^{\pm }\equiv g^{\pm }\exp \left( -\frac{y^{0}}{2a_{\#}}\right) \equiv
\left( g_{\pm }^{L}\right) ^{-1}\text{,} \\ 
g_{R}^{\pm }\equiv \overline{g}^{\pm }\exp \left( -\frac{y^{0}}{2a_{\#}}%
\right) \equiv \left( g_{\pm }^{R}\right) ^{-1}\text{.}%
\end{array}%
\end{equation*}%
The right-invariant spin connections are 
\begin{equation}
\begin{array}{c}
\Omega _{\pm }^{L}\equiv \left( \mathbf{d}g_{L}^{\pm }\right) g_{\mp }^{L}%
\text{;} \\ 
\Omega _{\pm }^{R}\equiv \left( \mathbf{d}g_{R}^{\pm }\right) g_{\mp }^{R}%
\text{.}%
\end{array}
\label{8}
\end{equation}

\emph{Effective} spin connections (\ref{3}) are formed by \emph{differentials%
} of each spinor field, when multiplied by its $PTC$ conjugate spinor:%
\begin{equation}
\begin{array}{c}
\psi ^{I}\mathbf{d}\psi _{I}=\mathbf{\psi }^{I}g_{I}\mathbf{d}\left( g^{I}%
\mathbf{\psi }_{I}\right) =\mathbf{\psi }^{I}\Omega _{I}\mathbf{\psi }_{I}%
\text{;} \\ 
\Omega _{I}\equiv \frac{i}{2}\mathbf{d}\zeta _{I}\equiv \frac{i}{2}\mathbf{d}%
\zeta _{I}^{\alpha }\sigma _{\alpha }=\frac{i}{2}\mathbf{d}\theta
_{I}^{\alpha }-\frac{1}{2}\mathbf{d}\varphi _{I}^{\alpha }\text{.}%
\end{array}
\label{9}
\end{equation}

The \emph{anti-Hermitian} ($aH$) parts $\frac{i}{2}\mathbf{d}\theta ^{\alpha
}\sigma _{\alpha }$ of the spin connections are the $u\left( 1\right) \times
su\left( 2\right) $ \emph{electroweak vector potentials}. The \emph{Hermitian%
} ($H$) parts $-\frac{1}{2}\mathbf{d}\varphi ^{\alpha }\sigma _{\alpha }$
are the \emph{gravitational} potentials. These measure the local \emph{%
dilation-boost flow} $\mathbf{d}\varphi ^{\alpha }\left( x\right) $ of the
spinfluid \cite{im}, \cite{xueg}. The path dependences, or \emph{holonomies,}
\begin{equation}
g^{I}\mathbf{dd}g_{I}\equiv K_{I}=\mathbf{d}\Omega _{I}+\Omega _{I}\wedge
\Omega _{I}\text{,}  \label{10}
\end{equation}%
of the spin connections are the \emph{spin curvatures,} or \emph{fields}.

The dilation/boost flow acts on the position quaternion, (\ref{4}) to
produce the \emph{vacuum} \emph{energy-momentum} distribution%
\begin{equation*}
\exp \left[ \frac{i}{a_{\#}}z^{\alpha }\left( x\right) \sigma _{\alpha }%
\right] \equiv \left( iq+p\right) \left( x\right) \in \mathbb{CM}%
_{\#}\subset T^{\ast }\mathbb{M}_{\#}\text{.}
\end{equation*}%
This assigns a \emph{position/momentum quaternion} $\left( iq+p\right)
\left( x\right) $ in the \emph{phase space} $T^{\ast }\mathbb{M}_{\#}$ to
each regular point $x\in \mathbb{M}_{\#}$ on the base space. The \emph{%
complex structure} on the complex quaternionic phase space $\mathbb{CM}_{\#}$
gives rise to the \emph{symplectic structure} of particle orbits on $T^{\ast
}\mathbb{M}_{\#}$ \cite{ahs}.

The dilation parameter%
\begin{equation*}
\varphi ^{0}\left( x\right) =a_{\#}^{-1}y^{0}\left( x\right)
\end{equation*}%
encodes local concentrations of rest energy, or \emph{mass.} Unlike the
boost current $\mathbf{d}\varphi ^{j}\left( x\right) $, the local energy
current $\mathbf{d}\varphi ^{0}\left( x\right) $ is not directly visible to
us as dwellers in a spacelike (constant $y^{0}$) cross section, because it
denotes the component $p^{0}$ of the 4-momentum flux \emph{normal} to our
expanding space $S_{3}\left( t\right) $.

However, if the phase flow 
\begin{equation}
\mathbf{d}\varsigma ^{\alpha }\equiv \frac{\partial \zeta ^{\alpha }}{%
\partial z^{\beta }}\mathbf{d}z^{\beta }+\frac{\partial \zeta ^{\alpha }}{%
\partial \bar{z}^{\beta }}\mathbf{d}\bar{z}^{\beta }  \label{11}
\end{equation}%
were \emph{analytic,} it would obey the Cauchy-Riemann equations: 
\begin{equation}
\frac{\partial \varsigma ^{\alpha }}{\partial \bar{z}^{\beta }}%
=0\Longrightarrow -\frac{\partial \theta ^{\alpha }}{\partial x^{\beta }}=%
\frac{\partial \varphi ^{\alpha }}{\partial y^{\beta }}\text{;\quad }\frac{%
\partial \theta ^{\alpha }}{\partial y^{\beta }}=\frac{\partial \varphi
^{\alpha }}{\partial x^{\beta }}\text{.}  \label{12}
\end{equation}%
We could then detect the dilation current, or \emph{rest energy,} by the 
\emph{frequency} 
\begin{equation}
-\frac{\partial \theta ^{0}}{\partial x^{0}}=\frac{\partial \varphi ^{0}}{%
\partial y^{0}}=\frac{m}{\text{%
h{\hskip-.2em}\llap{\protect\rule[1.1ex]{.325em}{.1ex}}{\hskip.2em}%
}}  \label{13}
\end{equation}%
of the matter wave, $\psi $. But \emph{energy} is the Noether charge under
Minkowsky-time translation. For the Dirac Lagrangian, $\mathcal{L}_{D}$ \cite%
{spingeo}, \cite{xueg}, this\emph{\ }turns out to be Planck's constant times
the frequency (\ref{13}), which we \emph{can} detect:%
\begin{equation}
\int_{B_{3}}\frac{\partial \mathcal{L}_{D}}{\partial \left( \partial
_{0}\psi _{I}\right) }\left[ \frac{\partial \psi _{I}}{\partial x^{0}}\right]
d^{3}v=\text{%
h{\hskip-.2em}\llap{\protect\rule[1.1ex]{.325em}{.1ex}}{\hskip.2em}%
}\int_{B_{3}}-\left( \frac{\partial \theta ^{0}\left( x\right) }{\partial
x^{0}}\right) e^{1}\wedge e^{2}\wedge e^{3}\text{.}  \label{14}
\end{equation}

\section{Dirac Spinors and Clifford Vectors}

Covariance of the Dirac equations on a curved, expanding spacetime $\mathbb{M%
}$ \cite{davies} rests on the local \emph{spinorization maps} 
\begin{equation}
\begin{array}{c}
S\equiv q_{\alpha }\left( x\right) E^{\alpha }\left( x\right) :E_{\beta
}\left( x\right) \longrightarrow q_{\beta }\left( x\right) \text{;} \\ 
\bar{S}\equiv \bar{q}_{\alpha }\left( x\right) E^{\alpha }\left( x\right)
:E_{\beta }\left( x\right) \longrightarrow \bar{q}_{\beta }\left( x\right) 
\text{.}%
\end{array}
\label{15}
\end{equation}%
These assign the fields; $q_{\beta }\left( x\right) \in gl\left( 2\mathbb{C}%
\right) _{L}$, $\bar{q}_{\beta }\left( x\right) \in gl\left( 2\mathbb{C}%
\right) _{R}$, of \emph{moving tetrads}---complex quaternions---to each
intrinsic spacetime increment $E_{\alpha }\left( x\right) \in T\mathbb{M}$.
The tetrads obey the \emph{Clifford algebra} of $\mathbb{M}$: 
\begin{equation}
\left[ q_{\alpha }\bar{q}_{\beta }+q_{\beta }\bar{q}_{\alpha }\right] \left(
x\right) =2g_{\alpha \beta }\left( x\right) \sigma _{0}\text{,}  \label{16}
\end{equation}%
where $g_{\alpha \beta }\left( x\right) $ is the \emph{metric tensor}, which
gives the \emph{scalar products} of Clifford tetrads, pairwise. The overbar
denotes \emph{quaternionic} conjugation---space ($P$) reversal.

The Clifford tetrads are sums of \emph{null tetrads:} tensor products of
some fundamental $L$- and $R$-chirality physical fields---the inertial
spinor fields, or \emph{vacuum spinors} (VS) \cite{penrose}, \cite{vdw}: the
covariantly constant (null) modes of the Dirac operators. The local tetrads
are \emph{defined} as $gl\left( 2,\mathbb{C}\right) $ matrices, with respect
to the vacuum spinors as bases: 
\begin{equation}
\begin{array}{c}
q_{\alpha }\left( x\right) =\sigma _{\alpha }^{\;A\dot{B}}\ell _{A}\left(
x\right) \otimes r_{\dot{B}}^{\;T}\left( x\right) \text{,\qquad }\bar{q}%
_{\alpha }=\sigma _{\alpha }^{\;\dot{U}V}r_{\dot{U}}\left( x\right) \otimes
\ell _{V}^{\;T}\left( x\right) \text{,}%
\end{array}
\label{17}
\end{equation}%
(sum on $A$, $\overset{\cdot }{B}=1,2$).

On an expanding, \emph{homogeneous} Friedmann 3-sphere $\left( T,\mathbb{S}%
_{3}\left( T\right) \right) $ with dilation parameter $y^{0}\equiv T$, the
normalized vacuum spinors and cospinors are%
\begin{equation}
\begin{array}{c}
\ell ^{\pm }\equiv \gamma ^{-\frac{1}{2}}\widehat{\ell }^{\pm }\equiv \left[
\ell _{1}^{+},\ell _{2}^{-}\right] \left( x\right) =\sigma _{0}\exp \frac{i}{%
2a_{\#}}\left[ \left( \pm x^{0}+iy^{0}\right) \sigma _{0}+x^{j}\sigma _{j}%
\right] \text{,} \\ 
r^{\mp }\equiv \gamma ^{-\frac{1}{2}}\widehat{r}^{\mp }\equiv \left[ r_{\dot{%
1}}^{-},r_{\dot{2}}^{+}\right] \left( x\right) =\sigma _{0}\exp \frac{i}{%
2a_{\#}}\left[ \left( \mp x^{0}+iy^{0}\right) \sigma _{0}+x^{j}\overline{%
\sigma }_{j}\right] \text{,} \\ 
r_{\pm }\equiv \gamma ^{-\frac{1}{2}}\widehat{r}_{\pm }\equiv \left[ r_{+}^{%
\dot{1}},r_{-}^{\dot{2}}\right] ^{T}\left( x\right) =\exp \frac{i}{2a_{\#}}%
\left[ \left( \pm x^{0}+iy^{0}\right) \sigma _{0}+x^{j}\overline{\sigma }_{j}%
\right] \sigma _{0}\text{,} \\ 
\ell _{\mp }\equiv \gamma ^{-\frac{1}{2}}\widehat{\ell }_{\mp }\equiv \left[
\ell _{-}^{1},\ell _{+}^{2}\right] ^{T}\left( x\right) =\exp \frac{i}{2a_{\#}%
}\left[ \left( \mp x^{0}+iy^{0}\right) \sigma _{0}+x^{j}\sigma _{j}\right]
\sigma _{0}\text{.}%
\end{array}
\label{18}
\end{equation}%
The $\left( +\right) $ charge index indicates propagation of the $u\left(
1\right) $ phase \emph{outward} (in the direction of cosmic expansion), $%
\frac{\partial \theta ^{0}}{\partial T}>0$; a $\left( -\right) $ charge
index indicates \emph{inward} propagation. For the vacuum spinors, charge is
coupled to spin \cite{spingeo}. We have indicated their charges in the
opposite position to their spin indices in (\ref{18}); then suppressed spin
indices by writing the moving frames for spinors and cospinors as $GL\left(
2,\mathbb{C}\right) $ matrices columnwise and row-wise respectively. $PTC$%
-reversal of a spinor produces the \emph{dual} cospinor, an \emph{equivalent}
representation of the $E$ group \cite{spingeo}. Thus there are $\mathbf{4}$
fundamental $E$ representations:%
\begin{equation*}
\begin{array}{ccc}
\ell ^{+}\sim r_{-} &  & r^{+}\sim \ell _{-} \\ 
\ell ^{-}\sim r_{+} &  & r^{-}\sim \ell _{+}\text{.}%
\end{array}%
\end{equation*}

The vacuum spinors in (\ref{18}) have \emph{conformal weights} $\gamma ^{-%
\frac{1}{2}}$, where $\gamma $ is the scale factor (\ref{5}) of our
expanding spatial hypersurface $S_{3}\left( T\right) $. This assures that
the covariant tetrads (\ref{17}), metric tensor (\ref{16}), and the
intrinsic volume element all scale properly with $\gamma $:%
\begin{equation}
\begin{array}{c}
\left\vert q_{\alpha }\right\vert \sim \gamma ^{-1}\sim \left\vert \overline{%
q}_{\beta }\right\vert :g_{\alpha \beta }\sim \gamma ^{-2} \\ 
\left\vert E^{\alpha }\right\vert \sim \gamma ^{+1}:g^{\alpha \beta }\sim
\gamma ^{+2}\text{;} \\ 
e^{0}\wedge e^{1}\wedge e^{2}\wedge e^{3}=\left\vert g\right\vert ^{\frac{1}{%
2}}E^{0}\wedge E^{1}\wedge E^{2}\wedge E^{3}=\gamma ^{-4}E^{0}\wedge
E^{1}\wedge E^{2}\wedge E^{3}\text{.}%
\end{array}
\label{19}
\end{equation}%
The vacuum intensity $\ell ^{\pm }r_{\mp }$ scales as $\gamma ^{-1}$ with
cosmic expansion; the intensity of the vacuum spinors at $T=0$, on $\mathbb{M%
}_{\#}$, has been normalized to $1$ (c.f. \cite{spingeo}). The \emph{vacuum
amplitude} scales as $\gamma ^{-\frac{1}{2}}=\exp \left[ -\frac{T}{2a_{\#}}%
\right] $, so the vacuum energy density $\gamma ^{-4}$ integrates to $1$, a 
\emph{constant}, in the dilated frame (\ref{19}).

We denote the bundle of vacuum spinors over $\mathbb{M}_{\#}$ as perturbed,
pathwise, by local sources as%
\begin{equation}
\begin{array}{c}
\ell ^{\pm }\left( x\right) \equiv \gamma ^{-\frac{1}{2}}\widehat{\ell }%
^{\pm }\exp \frac{i}{2}\left[ \theta _{L}^{\alpha \pm }\left( x\right)
+i\varphi _{L}^{\alpha \pm }\left( x\right) \sigma _{\alpha }\right] \\ 
r^{\pm }\left( x\right) \equiv \gamma ^{-\frac{1}{2}}\widehat{r}^{\pm }\exp 
\frac{i}{2}\left[ \theta _{R}^{\alpha \pm }\left( x\right) +i\varphi
_{R}^{\alpha \pm }\left( x\right) \overline{\sigma }_{\alpha }\right]%
\end{array}
\label{20}
\end{equation}%
Tensor products of vacuum spinors make the \emph{null tetrads.} Sums and
differences of these make the ($spin$-$1$) tetrads (\ref{17}). Products of
the tetrads make the ($spin$-$2$) metric tensor (\ref{16}).

It is a remarkable fact \cite{im} that (up to a global $GL\left( 2,\mathbb{C}%
\right) $ transformation) the \emph{same} perturbed vacuum spinor fields (%
\ref{20}) that factor the Clifford tetrads must be used as \emph{moving spin
frames} for the ($spin$-$\frac{1}{2}$) \emph{matter fields} \cite{im}, \cite%
{sachs}, \cite{vdw}, in order to preserve covariance of the Dirac equations
in curved spacetime!

This is the principle of \textquotedblleft general covariance", or

\begin{quote}
\emph{Spin Principle, }$\mathbf{S}$\emph{:} \emph{Spinor fields} are \emph{%
physical}. All matter and gauge fields are (sums of) tensor products of
spinor fields and their differentials \cite{spingeo}, \cite{penrose}, \cite%
{vdw}, \cite{keller}.
\end{quote}

Spinors come in $2^{3}=8$ varieties: \emph{left} or \emph{right-chirality}
(handed-nes, i.e. $su\left( 2\right) $ phase rotation with spatial
translation),\emph{\ heavy} or \emph{light temporality} (outward or inward
propagation in cosmic time, $T$), and positive or negative \emph{charge} ($%
u\left( 1\right) $ phase shift with cosmic time).

Spinors are real \cite{spingeo}. Material particles are localized
configurations of spinor fields; spacetime is the homgeneous distribution of
the vacuum spinors between particles. Our moving spacetime tetrads $%
E_{\alpha }\left( x\right) \in T\mathbb{M}\left( x\right) $ \emph{are} the
inverse images under spinorization maps (\ref{15}) of the Clifford tetrads $%
q_{\alpha }\left( x\right) $ and $\bar{q}_{\alpha }\left( x\right) $:
standing-wave distributions (\ref{17}) of internal $gl\left( 2,\mathbb{C}%
\right) $ phase stepped off by the vacuum spinors.

Spinorization maps $S$ and $\bar{S}$ are implemented \emph{physically} by
the \emph{spin} connections 
\begin{equation}
\begin{array}{c}
\Omega _{L}\left( x\right) \equiv g^{L}\mathbf{d}g_{L}=\frac{i}{2}\left[
a_{\#}^{-1}\sigma _{\alpha }\left( x\right) +W_{\alpha }\left( x\right) %
\right] e^{\alpha } \\ 
=\frac{i}{2}\left[ a_{\#}^{-1}q_{\alpha }\left( x\right) +w_{\alpha }\left(
x\right) \right] E^{\alpha }\text{,} \\ 
\\ 
\Omega _{R}\left( x\right) \equiv g^{R}\mathbf{d}g_{R}=\frac{i}{2}\left[
a_{\#}^{-1}\overline{\sigma }_{\alpha }\left( x\right) +\overline{W}_{\alpha
}\left( x\right) \right] e^{\alpha } \\ 
=\frac{i}{2}\left[ a_{\#}^{-1}q_{\alpha }\left( x\right) +\overline{w}%
_{\alpha }\left( x\right) \right] E^{\alpha }\text{,}%
\end{array}
\label{21}
\end{equation}%
the \emph{left invariant Maurer-Cartan forms}, given first in the fiducial $%
\mathbb{M}_{\#}$ reference frame, and then in the dilated (but static) frame%
\begin{equation}
\begin{array}{c}
E^{\alpha }=\gamma e^{\alpha }\text{;} \\ 
q_{\alpha }=\gamma ^{-1}\sigma _{\alpha }\text{.}%
\end{array}
\label{22}
\end{equation}%
In an intrinsic reference frame co-expanding with the Friedmann flow, the
temporal spin connections%
\begin{equation*}
\Omega _{0}^{\pm }=\frac{i}{2a_{\#}}\left( \pm i+i\overset{\cdot }{y}%
_{0}\right)
\end{equation*}%
pick up negative \textquotedblleft kinetic energy" terms in $\frac{\overset{%
\cdot }{y}^{0}}{2a_{\#}}=\frac{\overset{\cdot }{\gamma }}{2}$, the \emph{rate%
} of cosmic expansion. This might be thought of as a sort of
\textquotedblleft Doppler shift" of energies in our expanding frame relative
to the static reference frame (\ref{22}) on $\mathbb{R}_{+}\times \mathbb{S}%
_{3}\left( a\right) $.

\section{The Topological Lagrangian and its Spinor Factorization}

In the $PT$-symmetric (gravitational) case, the electroweak vector
potentials $w_{\alpha }$ and $\overline{w}_{\alpha }$ in (\ref{21}) vanish,
and the spin connections \emph{are} the tetrads. The exterior product of all 
$4$ spin connections is a natural \emph{topological Lagrangian}, whose
action is the covering number of spin space over spacetime.

Products of cospinors and spinor differentials make \emph{effective spin
connections:} Lie-algebra valued $1$ forms (\ref{3}) that give the internal
chiral $gl\left( 2,\mathbb{C}\right) $ phase increments stepped off over
each infinitesimal spacetime displacement.

Now it takes tensor products of $\mathbf{4}$ pairs of cospinors and spinor
differentials to make a \emph{natural} $4$ form---e.g. a Lagrangian density $%
4$ form%
\begin{equation}
\mathcal{L}_{g}\in \Lambda ^{4}\subset \otimes ^{8}  \label{23}
\end{equation}%
invariant under the group $E_{P}$ of passive spin isometries in curved
spacetime. In the $PTC$-symmetric case, the wedge product of all four of the
Hermitian spin connections (\ref{21}) makes the invariant $4$\emph{-volume}
element---a Clifford scalar \cite{spingeo}: 
\begin{equation}
\begin{array}{c}
\frac{i}{2}Tr\Omega ^{R}\wedge \Omega _{L}\wedge \Omega ^{L}\wedge \Omega
_{R}=\left( \frac{1}{16a_{\#}^{4}}\right) \sigma _{0}e^{0}\wedge e^{1}\wedge
e^{2}\wedge e^{3}\equiv \left( \frac{1}{16a_{\#}^{4}}\right) \sigma _{0}%
\mathbf{d}^{4}v \\ 
=\left( \frac{1}{16a_{\#}^{4}}\right) \left\vert g\right\vert ^{\frac{1}{2}%
}\sigma _{0}E^{0}\wedge E^{1}\wedge E^{2}\wedge E^{3}\equiv \left( \frac{1}{%
16a_{\#}^{4}}\right) \left\vert g\right\vert ^{\frac{1}{2}}\sigma _{0}%
\mathbf{d}^{4}V\text{.}%
\end{array}
\label{24}
\end{equation}%
Here $\left\vert g\right\vert ^{\frac{1}{2}}$ is the square root of (minus)
the determinant of the covariant metric tensor $g_{\alpha \beta }$ of (\ref%
{16}): the inverse expansion factor of a $4$-volume element on $\mathbb{M}$
comoving with Friedmann flow\footnote{%
For a \emph{static} frame $E^{\alpha }=\gamma e^{\alpha }$, $\left\vert
g\right\vert ^{\frac{1}{2}}=\gamma ^{-4}$. For a Lorenz frame expanding at
rate $\overset{\cdot }{y}^{0}=a_{\#}\overset{\cdot }{\gamma }$, the comoving 
$1$ forms are $E^{\alpha \prime }=\left( 1+a_{\#}^{2}\overset{\cdot }{\gamma 
}^{2}\right) ^{\frac{1}{2}}E^{\alpha }$, and the inverse $4$-volume
expansion factor is%
\begin{equation*}
\left\vert g^{\prime }\right\vert ^{\frac{1}{2}}=\left( 1+a_{\#}^{2}\overset{%
\cdot }{\gamma }^{2}\right) ^{2}\left\vert g\right\vert ^{\frac{1}{2}}\equiv
\gamma ^{\prime -4}\left\vert g\right\vert ^{\frac{1}{2}}\text{.}
\end{equation*}%
} relative to the unit volume element on $\mathbb{M}_{\#}$.

The \emph{simplest} Lagrangian that is an $E_{P}$-invariant $4$ form is thus
the \emph{topological Lagrangian}%
\begin{equation}
\mathcal{L}_{T}\equiv \frac{i}{2}Tr\Omega ^{R}\wedge \Omega _{L}\wedge
\Omega ^{L}\wedge \Omega _{R}\text{;}  \label{25}
\end{equation}%
the \emph{Maurer-Cartan} $4$ form. The \emph{topological action} 
\begin{equation}
S_{T}\equiv \frac{i}{2}\int_{\widehat{\mathbb{M}}}Tr\Omega ^{R}\wedge \Omega
_{L}\wedge \Omega ^{L}\wedge \Omega _{R}=-16\pi ^{3}C_{2}  \label{26}
\end{equation}%
measures the \emph{covering number} (second Chern number) of spin space over
the \emph{regular region} \cite{tqac}, \cite{penrose}: the perturbed vacuum $%
\widehat{\mathbb{M}}\equiv \mathbb{M}_{\#}\backslash \cup D^{J}$ outside the
codimension-$J$ \emph{singular loci;} the supports $\gamma _{4-J}$ of
massive particles.

More generally, our base spacetime $\mathbb{M}_{\#}$ is \emph{stratified}
into the \emph{regular stratum} $D^{0}=\widehat{\mathbb{M}}\equiv \mathbb{M}%
_{\#}\backslash \cup D^{J}$, where geometrical optics ansatz (\ref{1}) holds
for all $4$ $PTC$-opposed pairs of spinor fields, and co-dimension-$J$ \emph{%
singular strata.} Here, the phases of $J=1,2,3,$ or $4$ pairs of spinors
either become singular, or break away from $PTC$-symmetry \cite{tqac}.

We call the union $\cup D^{\alpha }$ of all these strata and their incidence
relations, together with the periods of all 1, 2, 3, and 4 forms quantized
over embedded homology cycles, the \emph{Spin}$^{c}$\emph{-4 complex,} $S$.

Each singular locus $D^{J}\subset S$ carries its own topological charges:
integrals of $J$ forms or their dual ($4-J$) forms over $J$ cycles or their
transverse ($4-J$) cycles \cite{tqac}. Each charge is quantized in integer
units. The spin connections (\ref{7}) derived from the canonical maps (\ref%
{6}) of spin space over spacetime correspond to covering number $C_{2}=1$.

The topological action $S_{T}$ is a \emph{conformal invariant} that comes in
topologically-quantized units \cite{tqac}. Any (local) rescaling of a
solution preserves $S_{T},$and so is still a solution. Thus, topological
Lagrangian $\mathcal{L}_{T}$ cannot be the Lagrangian for massive fields 
\emph{inside} their worldtubes, because \emph{mass} arises from the breaking
of scale invariance.

Elsewhere \cite{capp}, \cite{spingeo} we exhibited a \textquotedblleft
grandparent\textquotedblright\ Lagrangian density for both the \emph{outer }%
(regular) region \cite{tqac}, where it reduces to $\mathcal{L}_{T}$, and the 
\emph{inner} (singular) region, inside the worldtubes of massive particles: 
\begin{equation}
\mathcal{L}_{g}\equiv i\mathbf{d}\psi ^{R\pm }\psi _{L\mp }\wedge \psi
^{L\pm }\mathbf{d}\psi _{R\mp }\wedge \mathbf{d}\psi ^{L\pm }\psi _{R\mp
}\wedge \psi ^{R\pm }\mathbf{d}\psi _{L\mp }  \label{27}
\end{equation}%
(average over all neutral sign combinations in which each spinor, or its
differential, appears exactly \emph{once}). $\mathcal{L}_{g}$ is the $%
\mathbf{8}$\emph{-spinor factorization} of the \emph{Maurer-Cartan }$4$\emph{%
\ form}.

One remarkable feature of $\mathcal{L}_{g}$ is \cite{spingeo}, \cite{xueg}
that it yields effective electroweak, strong, and gravitational \emph{field
actions} in $\widehat{\mathbb{M}}$ when each field is expanded as the sum of
a vacuum (dark energy) distribution with vacuum amplitudes of order $\gamma
^{-\frac{1}{2}}$, and a \textquotedblleft broken out\textquotedblright\
matter spinor: 
\begin{equation}
\begin{array}{ccc}
\psi _{L_{\pm }}\equiv \gamma ^{-\frac{1}{2}}\widehat{\ell }_{\pm }+\xi
_{\pm }\text{,} &  & \psi _{R_{\pm }}\equiv \gamma ^{-\frac{1}{2}}\widehat{r}%
_{\pm }+\eta _{\pm }\text{,} \\ 
\psi ^{L_{\pm }}\equiv \gamma ^{-\frac{1}{2}}\widehat{\ell }^{\pm }+\chi
^{\pm }\text{,} &  & \psi ^{R_{\pm }}\equiv \gamma ^{-\frac{1}{2}}\widehat{r}%
^{\pm }+\zeta ^{\pm }\text{.}%
\end{array}
\label{28}
\end{equation}

All $\mathbf{8}$ fields in $\mathcal{L}_{g}$ may be varied independently. 
\emph{Outside} the worldtubes $D_{J}$ of massive particles, the action $%
S_{g} $ is stationarized by the \emph{outer solution.} This turns out \cite%
{spingeo}, \cite{im} to be both $C$ and $PT$-symmetric: it preserves the
\textquotedblleft inner" products $\psi ^{R_{\pm }}\psi _{L_{\mp }}\ $of $%
PTC $-conjugate spinors as conformal invariants \cite{spingeo}. Under $PTC$
symmetry \cite{spingeo} in $\widehat{\mathbb{M}}$, outside the worldtubes of
massive particles, we may use the simpler form%
\begin{equation}
\mathcal{L}_{g}\overset{PTC}{\Longrightarrow }i\gamma ^{-2}\mathbf{d}\psi
^{R\pm }\wedge \mathbf{d}\psi _{R\mp }\wedge \mathbf{d}\psi ^{L\pm }\wedge 
\mathbf{d}\psi _{L\mp }  \label{29}
\end{equation}%
for the field Lagrangian.

What happens \emph{inside} the worldtubes $B_{4}$ of massive particles \cite%
{spingeo} is that Left- and Right-chirality of matter spinors, which have
opposite nonAbelian magnetic charges, \emph{bind} to form localized $PT$%
-antisymmetric \emph{bispinor particles} like 
\begin{equation}
e_{-}\equiv \left( \xi _{-}\left( x\right) \oplus \eta _{-}\left( x\right)
\right) \text{,}  \label{30}
\end{equation}%
the \emph{electron.}

\section{Vacuum Spinors and Dirac Mass}

Mass arises from the breaking of scale invariance. The mechanism that breaks
scale invariance and endows bispinor particles with \emph{inertial mass}
emerges in a remarkable way \cite{spingeo}, \cite{im} when ansatz (\ref{28})
for the perturbed spinor fields is inserted into Lagrangian (\ref{27}). We
summarize below (please see Appendix also).

Massless fields, like the vacuum spinors in (\ref{28}), contribute terms
with conformal weight $\gamma ^{-4}$ to the Lagrangian. These integrate to
terms of $O\left( \gamma ^{0}\right) $---constant terms---like homogeneous
the vacuum action of $-16\pi ^{3}W$ (\ref{26}).

As the scale factor $\gamma $ increases past $1$, terms of $O\left( \gamma
^{-3}\right) $ in $\mathcal{L}_{g}$ localize about 3 cycles, corresponding
to the condensation of bispinor particles (leptons) from \textquotedblleft
seed" perturbations wrapped about $3$ cycles in the perturbed vacuum. It is
these $O\left( \gamma ^{-3}\right) $ terms that give the effective \emph{%
massive Dirac action} \cite{spingeo}. Heuristically, what happens is this.

Inside the worldtube $B_{4}$ of the massive bispinor particle $e_{-}$, the $%
PTC$-opposed pairs of matter fields $\left( \xi _{-},\zeta ^{+}\right) $ and 
$\left( \eta _{-},\chi ^{+}\right) $ of (\ref{28}) undergo \emph{mass
scatterings }\cite{penrose}: \emph{Spin}$^{c}$-$4$ resonances with the
remaining vacuum fields and differentials, which pair to form effective spin
connections (\ref{9}): 
\begin{equation*}
\begin{array}{c}
i\hat{\Omega}\wedge \hat{\Omega}\wedge \hat{\Omega}\wedge \zeta ^{+}\mathbf{d%
}\xi _{-} \\ 
i\left( \frac{i}{2a_{\#}}\right) ^{3}q_{1}\overline{q}_{2}\overline{q}%
_{3}E^{1}\wedge E^{2}\wedge E^{3}\wedge \zeta ^{+}D_{0}\xi _{-}E^{0} \\ 
=\left( \frac{1}{2a_{\#}}\right) ^{3}\gamma ^{-1}\sigma _{1}\gamma ^{-1}%
\overline{\sigma }_{2}\gamma ^{-1}\overline{\sigma }_{3}E^{1}\wedge
E^{2}\wedge E^{3}\wedge \zeta ^{+}D_{0}\xi _{-}E^{0} \\ 
=i\left( \frac{1}{2a_{\#}}\right) ^{3}\gamma ^{-3}\sigma _{0}\zeta
^{+}D_{0}\xi _{-}\mathbf{d}^{4}V\text{.}%
\end{array}%
\end{equation*}%
Here $D_{\alpha }\xi =\gamma ^{-1}\partial _{\alpha }\xi $ are covariant
derivatives (\ref{2}) in the \emph{dilated} coframe, $E^{\alpha }=\gamma
e^{\alpha }$, or $E^{\alpha \prime }=\gamma ^{\prime }e^{\alpha }$ in a 
\emph{coexpanding} coframe. Each $Spin^{c}$-4 resonance reconstructs a $3$%
\emph{-volume element} dual to the light spinor gradient, giving the
intrinsic Dirac operators $i\overline{\sigma }_{\alpha }\nabla _{\alpha }$
and $i\sigma _{\alpha }\overline{\nabla }_{\alpha }$ \cite{spingeo}.

Meanwhile, multilinear (tensor) products of $4$ vacuum differentials and $2$
vacuum spinors produce $spin$-1 tensors like%
\begin{equation}
\begin{array}{c}
\mathbf{d}\widehat{r}_{\pm }\widehat{\ell }^{\mp }\wedge \widehat{\ell }%
_{\pm }\mathbf{d}\widehat{r}^{\mp }\wedge \mathbf{d}\widehat{\ell }^{\mp
}\otimes \mathbf{d}\widehat{\ell }_{\pm }\equiv -\left( \frac{i}{2a_{\#}}%
\right) ^{4}\sigma _{0}\left( x\right) d^{4}V \\ 
\mathbf{d}\widehat{\ell }_{\pm }\widehat{r}^{\mp }\wedge \widehat{r}_{\pm }%
\mathbf{d}\widehat{\ell }^{\mp }\wedge \mathbf{d}\widehat{r}^{\mp }\otimes 
\mathbf{d}\widehat{r}_{\pm }\equiv \left( \frac{i}{2a_{\#}}\right) ^{4}%
\overline{\sigma }_{0}\left( x\right) d^{4}V\text{.}%
\end{array}
\label{31}
\end{equation}%
These couple pairs of light and heavy matter spinors, contributing the
effective \emph{mass} term, 
\begin{equation}
\mathcal{L}_{M}=-\left( \frac{i}{2a_{\#}}\right) ^{4}\left[ \zeta ^{+}\sigma
_{0}\left( x\right) \eta _{-}-\chi ^{+}\overline{\sigma }_{0}\left( x\right)
\xi _{-}\right] \gamma ^{-3}\mathbf{d}^{4}V\text{,}  \label{32}
\end{equation}%
to complete the Dirac Lagrangian $\mathcal{L}_{D}$ \cite{spingeo}.

Upon variation with respect to the heavy (light) spin vectors, $\mathcal{L}%
_{D}$ gives the \emph{massive Dirac equations }coupling the light (heavy)
envelopes of matter spinors \cite{spingeo}, written with respect to a moving
frame of unit spin matrices $\sigma ^{\alpha }\left( x\right) $, $\overline{%
\sigma }^{\alpha }\left( x\right) $ and covariant derivatives $\nabla
_{\alpha }\left( x\right) $, $\overline{\nabla }_{\alpha }\left( x\right) $
intrinsic to our dilated Friedmann $3$-brane $S_{3}\left( T\right) $:%
\begin{equation}
\begin{array}{c}
i\sigma ^{\alpha }\mathbf{\nabla }_{\alpha }\mathbf{\xi }_{-}=\frac{\beta
^{-1}}{2a_{\#}}\mathbf{\eta }_{-}\text{,\qquad }i\left( \mathbf{\nabla }%
_{\alpha }\mathbf{\chi }^{+}\right) \sigma ^{\alpha }=\frac{\beta }{2a_{\#}}%
\mathbf{\zeta }^{+} \\ 
i\overline{\sigma }^{\alpha }\overline{\mathbf{\nabla }}_{\alpha }\mathbf{%
\eta }_{-}=\frac{\beta ^{-1}}{2a_{\#}}\mathbf{\xi }_{-}\text{,\qquad }%
i\left( \overline{\mathbf{\nabla }}_{\alpha }\mathbf{\zeta }^{+}\right) 
\overline{\sigma }^{\alpha }=\frac{\beta }{2a_{\#}}\mathbf{\chi }^{+}\text{.}%
\end{array}
\label{33}
\end{equation}

These govern the lightest perturbations that unfold from the conformal
vacuum as $\gamma $ exceeds $1$ (see Appendix). The electron mass \cite{im},%
\begin{equation}
m_{e}=\frac{\beta }{2a_{\#}}\text{,}  \label{34}
\end{equation}%
turns out to be the inverse of the equilibrium diameter $2a_{\#}$ of our
Friedmann $3$-brane \cite{im}, \cite{fried} in the $\mathbb{M}_{\#}$ ($\beta
=1$) reference frame. In an intrinsic frame coexpanding with the Friedmann
flow $\overset{\cdot }{y}^{0}\equiv \overset{\cdot }{T}$,%
\begin{equation*}
\beta =\left[ \frac{1+\overset{\cdot }{T}}{1-\overset{\cdot }{T}}\right] 
\text{.}
\end{equation*}

The mass (\ref{34}) of a \textquotedblleft free" electron is an example of 
\emph{quantization of action:} a $1$ form integrated over a $1$-cycle $%
\gamma _{1}\in \mathbb{M}_{\#}$ \emph{transverse} to the electron's $3$%
-support $\ast D^{1}=B_{3}$:%
\begin{equation*}
\doint\limits_{\gamma _{1}}Edt-\mathbf{p}d\mathbf{x}=\int_{0}^{2\pi
a_{\#}}m_{e}dt=\pi
\end{equation*}%
(where 
h{\hskip-.2em}\llap{\protect\rule[1.1ex]{.325em}{.1ex}}{\hskip.2em}%
$=1$).

The worldtube $B_{4}=B_{3}\times \gamma _{1}$ of a massive Dirac particle
shares a common boundary $\partial B_{4}\in \partial \widehat{\mathbb{M}}$
with the perturbed vacuum $\widehat{\mathbb{M}}$. On a microscopic scale,
(Section 8) it is \emph{mass scatterings} on $\partial B_{4}$---the \emph{%
discrete} form of $Spin^{c}$-4 resonances\ (\ref{32}) with the vacuum spinor
fields (\ref{31})---that channel the \textquotedblleft null
zig-zags\textquotedblright\ of the Dirac propagator \cite{ord}, \cite%
{penrose}, \cite{xueg} into a timelike worldtube $B_{4}$. Inside $B_{4}$,
the light spinors ($\xi _{-}$, $\eta _{-}$) zig zag forward (outward) in
cosmic time $T$, while the heavy spinors ($\chi ^{+}$, $\zeta ^{+}$) zig zag
backward (inward), producing a net drift that is only \emph{statistically}
forward in $T$. At each vertex where a light and heavy spinor meet, the 
\emph{photons} $\gamma ^{\circlearrowleft }\equiv \xi _{-}\otimes \zeta ^{+}$
or $\gamma ^{\circlearrowright }\equiv \eta _{-}\otimes \chi ^{+}$ appear to
be created or annihilated. It is the statistics of this process---the
stochastic interactions of four matter spinors with four vacuum
spinors---that yields the massive Dirac-Maxwell propagator of Q.E.D. \emph{%
exactly} \cite{ord}.

Meanwhile, it is easy to see why no massive particle can move faster than
light---which is \textquotedblleft all zig, and no zag": because its
internal zigs and zags, though all lightlike, propagate in opposite
directions!

The \emph{vacuum fields} create \emph{Dirac mass}---the resistance of a
bispinor particle to acceleration. Qualitatively, when a particle, $P$, is
accelerated by $\bigtriangleup v$ over one mass scattering time $%
\bigtriangleup t$, the vacuum spinors impart a greater momentum change $%
\bigtriangleup p\equiv p^{-}-p^{+}$ to the \textquotedblleft trailing
surface" $\partial B_{4}^{-}$ of its worldtube boundary than to its
\textquotedblleft leading surface" $\partial B_{4}^{+}$. The \emph{frequency}
of mass scatterings is (proportional to) the rest mass of $P$; $\left(
\bigtriangleup t\right) ^{-1}\sim m$. Thus $\bigtriangleup p\sim
m\bigtriangleup v$; the discrete version of Newtons law for inertial forces.
This is Mach's principle in action. Quantitatively, all of the relativistic
kinematics of massive particles may be derived from mass scatterings with
the vacuum spinor fields on the worldtube boundary \cite{preprint}.

Reciprocally, massive particles perturb the vacuum fields. On the
macroscopic level, \emph{gravitation} is the spacetime curvature on the
boundary of the world tube of a test particle caused by perturbations due to
sources. We derive Einstein's field equations in the next section by
matching the \emph{spin curvatures} due to the source to the \emph{%
stress-energy} on the moving boundary $\partial B_{4}$ of a test particle's
worldtube.

\section{Matching Boundary Vorticity to Energy Momentum Flux}

The form of the effective matter Lagrangian $4$ form, $\mathcal{L}_{M}$, 
\emph{inside} the worldtube $B_{4}$, and its stress-energy $3$ form $\ast T$%
, depends on the particle. But the form of the \emph{field} Lagrangian
outside the worldtubes is universal. This gives us just enough information
to match the integral of the \emph{outer,\ field} $3$ form $G$, and the 
\emph{inner,} \emph{energy-momentum} flux $\ast T$ of the matter fields on
the \emph{moving} \emph{boundary} $\partial B_{4}\left( \tau \right) $, and
thus \emph{derive} Einstein's field equations. We outline the steps below.
For convenience, we work in our fiducial reference frame $e^{\alpha }\in
T^{\ast }\mathbb{M}_{\#}$; then translate our results to a dilated frame $%
E^{\alpha }=\gamma e^{\alpha }\in T^{\ast }\mathbb{M}$.

\begin{enumerate}
\item Write the total action $S_{g}$ as the sum of the \emph{field} terms 
\emph{outside} $B_{4}\left( \tau \right) $ (in $\widehat{\mathbb{M}}\equiv 
\mathbb{M}_{\#}\backslash B_{4}$), and \emph{matter} terms \emph{inside}: 
\begin{equation}
S_{g}=\frac{i}{2}\int_{\widehat{\mathbb{M}}}Tr\Omega ^{R}\wedge \Omega
_{L}\wedge \Omega ^{L}\wedge \Omega _{R}+\int_{B_{4}\left( \tau \right) }%
\mathcal{L}_{M}\equiv S_{F}+S_{M}\text{.}  \label{35}
\end{equation}%
Here $\tau $ is a proper time parameter along the particles' world tubes,
projected to $\mathbb{M}_{\#}$ \cite{im}.

\item Transform the \emph{field} term via integration by parts using the
Bianchi identity 
\begin{equation*}
\mathbf{d}K=K\wedge \Omega -\Omega \wedge K\text{.}
\end{equation*}%
The result is \cite{spingeo}%
\begin{equation}
\begin{array}{c}
S_{F}=\frac{i}{2}\int_{\widehat{\mathbb{M}}}Tr\left[ K_{L}\wedge
K_{R}+G_{L}\wedge G_{R}\right] -Tr\left[ \Omega _{L}\wedge \left(
K_{L}+K_{R}\right) \wedge \Omega _{R}+P\right]  \\ 
-i\int_{\partial B_{4}\left( \tau \right) }Tr\left[ \Omega _{L}\wedge
K_{R}+K_{L}\wedge \Omega _{R}\right] \text{,}%
\end{array}
\label{36}
\end{equation}%
where $P$ means space reversal. The first term is the chiral version of the
action in the electroweak and strong fields \cite{spingeo}, \cite{xueg}.
These may be combined to make the net (left) spin curvature $2$ form%
\begin{equation}
H_{L}\equiv K_{L}\oplus G_{L}\equiv \left( K_{L\gamma \delta }^{0}\sigma
_{0}\oplus K_{L\gamma \delta }^{j}\sigma _{j}\oplus G_{L}^{jk}\sigma
_{j}\otimes \sigma _{k}\right) e^{\gamma }\wedge e^{\delta }  \label{37}
\end{equation}%
(and similarly for $H_{R}$, with the $\sigma _{\alpha }$ replaced with $%
\overline{\sigma }_{\alpha }$). $H_{L}$ and $H_{R}$ take their values in the 
\emph{tensor product} $gl\left( 2,\mathbb{C}\right) _{\uparrow }\otimes
gl\left( 2,\mathbb{C}\right) _{\downarrow }$ of \emph{heavy} (baryonic) and 
\emph{light} (leptonic) Lie algebras. This decomposes into the direct sum%
\begin{equation*}
gl\left( 2\right) _{\uparrow }\otimes gl\left( 2\right) _{\downarrow }=%
\mathbb{C}\left[ u\left( 1\right) \oplus \left( su\left( 2\right) _{\uparrow
}\oplus su\left( 2\right) _{\downarrow }\right) \oplus su\left( 3\right) %
\right] 
\end{equation*}%
of complexified $u\left( 1\right) $ (electromagnetic), $su\left( 2\right) $
(weak; baryonic $\oplus $ leptonic), and $su\left( 3\right) $ (strong) Lie
algebras \cite{spingeo}.
\end{enumerate}

The \emph{electroweak} fields come from the \emph{antiHermitian} parts $%
\frac{i}{2}d\theta ^{\alpha }\sigma _{\alpha }$ of the $gl\left( 2,\mathbb{C}%
\right) $-valued vector potentials \cite{penrose}. Their \emph{Hermitian}
parts $\frac{1}{2}d\varphi ^{\alpha }\sigma _{\alpha }$ give the second term
in (\ref{36}), which contains the \emph{Palatini action} for the \emph{%
gravitational field }$K\equiv K_{L}\oplus K_{R}$. \emph{Both} the Hermitian
and nonHermitian potentials contribute to strong ($G$) fields in (\ref{37}).
But it is the third term---the \emph{boundary integral} term in field action
(\ref{36})---that couples fields to source currents in the next steps.

\begin{enumerate}
\item[3.] Express the boundary integral in terms of the matrix-valued \emph{%
spacetime curvature} $2$ form \cite{mtw} 
\begin{equation}
\mathcal{R}_{\alpha }^{\;\beta }\equiv R_{\alpha \;\gamma \delta }^{\;\beta
}e^{\gamma }\wedge e^{\delta }\text{.}  \label{38}
\end{equation}%
$\mathcal{R}$ accepts the area element $\left[ e_{\gamma },e_{\delta }\right]
$ and returns the holonomy \emph{(rotation)} matrix around it, with matrix
elements $\mathcal{R}_{\alpha }^{\;\beta }$.

\item[4.] Rewrite all spacetime vectors as Clifford ($C$) vectors, using
spinorization maps (\ref{15}), with $q_{\alpha }=\sigma _{\alpha }$ on $T%
\mathbb{M}_{\#}$. Now re-express the $PT$-symmetric part of the spacetime
curvature matrix, acting on a basis $C$ vector, in terms of $C$ vectors
multiplying the $gl\left( 2,\mathbb{C}\right) $-valued \emph{spin-curvature} 
$2$ forms \cite{sachs}, \cite{xueg}: 
\begin{equation}
\mathcal{R}_{\alpha }^{\;\beta }\sigma _{\beta }=\sigma _{\alpha }K_{R}+K_{L}%
\overline{\sigma }_{\alpha }\text{,}  \label{39}
\end{equation}%
where%
\begin{equation}
\begin{array}{c}
K_{L}\equiv K_{L\;\beta \gamma }^{\;\alpha }\sigma _{\alpha }e^{\beta
}\wedge e^{\gamma }\text{,} \\ 
K_{R}\equiv K_{R\;\beta \gamma }^{\;\alpha }\overline{\sigma }_{\alpha
}e^{\beta }\wedge e^{\gamma }\text{.}%
\end{array}
\label{40}
\end{equation}

\item[5.] Using Cartan's $C$ vector-valued $1$ form, 
\begin{equation}
\mathbf{d}q\left( x\right) \equiv \mathbf{d}\left( \sigma _{\alpha
}x^{\alpha }\right) \equiv \sigma _{\alpha }e^{\alpha }\text{,}  \label{41}
\end{equation}%
recognize the trace of the $gl\left( 2,\mathbb{C}\right) $-valued $3$ form 
\begin{equation}
\begin{array}{c}
\left( \frac{1}{2a_{\#}}\right) G\equiv \frac{1}{2a_{\#}}\mathbf{d}q\wedge 
\mathcal{R}=-i\left[ \Omega _{L}\wedge K_{R}+K_{L}\wedge \Omega _{R}\right]
\\ 
\equiv \left( \frac{1}{2a_{\#}}\right) G_{\;\beta }^{\alpha }\sigma _{\alpha
}\epsilon _{\;\gamma \delta \mu }^{\beta }e^{\gamma }\wedge e^{\delta
}\wedge e^{\mu }%
\end{array}
\label{42}
\end{equation}

as the \emph{integrand} in the \emph{outer form} of the boundary integral in
(\ref{36}). The $C$-vector-valued $3$ form $G$ is Wheeler's \cite{mtw}
\textquotedblleft moment of rotation tensor\textquotedblright : the
\textquotedblleft vorticity"\emph{\ }$\mathcal{R}$ of the \emph{spinfluid} 
\emph{dilation-boost flow} times the normal \emph{moment arm} to the area
element.

\item[6.] The \emph{inner} form of the boundary integral is the
energy-momentum $C$ vector $\mathcal{P}\equiv \mathcal{P}^{\alpha }\sigma
_{\alpha }$ of the matter fields inside the worldtube $B_{4}$ of the moving
particle. Detect this by displacing $B_{4}$ by the spacetime increment $%
t\equiv \bigtriangleup x^{\alpha }$ and rewriting the change in the action
as the \emph{surface} integral of a $C$-vector-valued $3$ form \emph{flux,} $%
\ast T\equiv \ast T^{\alpha }\sigma _{\alpha }$, across the moving boundary 
\cite{neu}, \cite{landl}: 
\begin{equation}
\mathcal{P}^{\alpha }\left( t\right) \equiv \int_{\partial B_{4}\left(
t\right) }\ast T^{\alpha }\text{.}  \label{43}
\end{equation}%
Here 
\begin{equation}
\ast T^{\alpha }\equiv \left[ \left( \frac{\partial \mathcal{L}}{\partial
\left( \partial _{\alpha }\psi _{I}\right) }\right) \partial _{\beta }\psi
_{I}-\delta _{\beta }^{\alpha }\mathcal{L}\right] \ast e^{\beta }\equiv
T_{\;\beta }^{\alpha }\ast e^{\beta }\text{,}  \label{44}
\end{equation}%
where%
\begin{equation*}
\ast e^{\beta }\equiv \epsilon _{\;\gamma \delta \mu }^{\beta }e^{\gamma
}\wedge e^{\delta }\wedge e^{\mu }
\end{equation*}%
is the $3$ form Hodge dual to the unit $1$ form $e^{\beta }$.%
\begin{equation*}
\ast T\equiv T_{\beta }^{\alpha }\sigma _{\alpha }\ast e^{\beta }\in \Lambda
^{3}\left( \mathbb{M}_{\#}\right)
\end{equation*}%
is a $C$-vector valued $3$ form: the \emph{energy-momentum} $3$ form. $\ast
T^{\alpha }$ is the Noether current under translation in the $e_{\alpha }$
direction. Taking $t=\tau $, the proper time along a particles worldline, (%
\ref{43}) gives $\mathcal{P}^{0}\left( \tau \right) $: the \emph{energy}
contained in the particle's spatial support $B_{3}\left( \tau \right) $; a
Clifford \emph{scalar}. This is its \emph{rest mass}.%
\newline%
More generally, if $\mathcal{L}$ is $t$-translation invariant, the action
contained \emph{inside} the closed worldtube $B_{4}\equiv \mathbb{S}%
_{1}\left( t\right) \times B_{3}\subset \mathbb{M}_{\#}$ on compactified
Minkowsky space is 
\begin{equation}
\mathcal{S}_{M}\left( B_{4}\right) =\int_{\mathbb{S}_{1}}dt\int_{B_{3}\left(
t\right) }\ast T=2\pi a_{\#}\int_{B_{3}}\ast T.  \label{45}
\end{equation}

\item[7.] Finally, equate the inner and outer expressions (\ref{44}) and (%
\ref{42}) in the action integral over the moving boundary 
\begin{equation*}
\partial B_{4}\left( t\right) \equiv B_{3}\left( t\right) -B_{3}\left(
0\right) +S_{2}\times I\left( t\right) 
\end{equation*}%
of a section of the worldtube of a particle in an external field. In the
particles rest frame, which is \textquotedblleft freely falling" in the
external field, the momentum flux over the boundary $S_{2}\times I\left(
\tau \right) $ vanishes, and we obtain: 
\begin{equation}
\begin{array}{c}
\frac{1}{2a_{\#}}\int_{B_{3}\left( \tau \right) }G=2\pi
a_{\#}\int_{B_{3}\left( \tau \right) }\ast T\Longrightarrow G=4\pi
a_{\#}^{2}\ast T\text{;} \\ 
\text{i.e. }G_{\;\beta }^{\alpha }=4\pi a_{\#}^{2}T_{\;\beta }^{\alpha }%
\end{array}
\label{46}
\end{equation}%
componentwise, since both integrals must be $E_{P}$-invariant. These are 
\emph{Einstein's field equations} \cite{mtw} on $\mathbb{M}_{\#}$, with a
gravitational constant of 
\begin{equation}
\kappa =\frac{a_{\#}^{2}}{2}\text{,}  \label{47}
\end{equation}%
the mean squared radius of the equilibrium Friedmann solution.
\end{enumerate}

Finally, we translate equation (\ref{46}) to the dilated spacetime frame $%
E^{\alpha }=\gamma e^{\alpha }\in T^{\ast }\mathbb{M}$ and Clifford-algebra
frame $q_{\alpha }=\gamma ^{-1}\sigma _{\alpha }\in C\left( T\mathbb{M}%
\right) $. Note (\ref{39}), (\ref{40}), (\ref{41}), (\ref{42}) that $G$ is a
\textquotedblleft Clifford Trivector" \cite{mtw}; it contains products of $3$
$C$ vectors $q_{\alpha }q_{\beta }q_{\gamma }=\gamma ^{-3}\sigma _{\alpha
}\sigma _{\beta }\sigma _{\gamma }$ (since the spin curvature $2$ forms
contain products of 2 $C$ vectors). Thus,%
\begin{equation*}
G_{\;\beta }^{\alpha \prime }=\gamma ^{-3}G_{\;\beta }^{\alpha }
\end{equation*}%
on $\mathbb{M}$. Since $\ast T\equiv T_{\;\beta }^{\alpha }\sigma _{\alpha
}=T_{\;\beta }^{\alpha \prime }\gamma q_{\alpha }$ is an ordinary $C$ vector,%
\begin{equation*}
T_{\;\beta }^{\alpha \prime }=\gamma ^{-1}T_{\;\beta }^{\alpha }
\end{equation*}%
on $\mathbb{M}$. Thus,%
\begin{equation}
G_{\;\beta }^{\alpha \prime }=\gamma ^{-2}4\pi a_{\#}^{2}T_{\;\beta
}^{\alpha \prime }\text{;}  \label{48}
\end{equation}%
Einstein's field equations on $\mathbb{M}$, with a gravitational constant of%
\begin{equation}
\kappa ^{\prime }=\gamma ^{-2}\frac{a_{\#}^{2}}{2}  \label{49}
\end{equation}%
in a dilated, but static-frame. In a frame comoving with cosmic expansion, $%
\gamma $ is replaced by%
\begin{equation*}
\gamma ^{\prime }\equiv \left( 1+a_{\#}^{2}\overset{\cdot }{\gamma }%
^{2}\right) ^{\frac{1}{2}}\gamma \text{.}
\end{equation*}

Note that the gravitational constant \emph{decreases} with radius $a\left(
t\right) =\gamma \left( t\right) a_{\#}$ of our Friedmann $3$-surface, and
with the Hubble constant $\frac{\overset{\cdot }{a}}{a}\left( t\right) =%
\frac{\overset{\cdot }{\gamma }}{\gamma }\left( t\right) $. Perhaps such
effects could be detected in astronomical data.

An independent check on our value (\ref{47}) for $\kappa $ on $\mathbb{M}%
_{\#}$ is provided by balancing $\ast T$ on the inside versus $G$ on the
outside of $B_{3}\equiv \mathbb{S}_{3}\left( a_{\#}\right) $, the \emph{%
equilibrium} Friedmann $3$-sphere. Here the curvature $2$ form and Cartan
unit $1$ form are%
\begin{equation*}
\mathcal{R}_{\;j}^{i}=\delta _{\;\ell }^{i}\delta _{jp}e^{\ell }\wedge e^{p}%
\text{;\quad }\mathbf{d}q\equiv \mathbf{d}\left( a_{\#}\exp \frac{i}{2a_{\#}}%
x^{j}\sigma _{j}\right) \text{.}
\end{equation*}%
This gives the moment-of-rotation $3$ form (\ref{42})%
\begin{equation*}
G=a_{\#}\epsilon _{jk\ell }^{\;n}\exp \left( \frac{i}{2a_{\#}}x^{m}\sigma
_{m}\right) \sigma _{n}e^{j}\wedge e^{k}\wedge e^{\ell }\text{;}
\end{equation*}%
the area $3$ form on $\mathbb{S}_{3}\left( a_{\#}\right) $ times the \emph{%
normal} $C$ vector.

Meanwhile, the $C$-vector-valued $3$ form $\ast T$ must integrate to a
constant---the rest energy in $\mathbb{S}_{3}\left( a_{\#}\right) $. $\ast T$
must be proportional to $a_{\#}^{-3}$ and, like $G$, be Clifford \emph{normal%
} to $\mathbb{S}_{3}\left( a_{\#}\right) $. We thus obtain relations (\ref%
{46}) and (\ref{47}) directly, by balancing the outward pressure $\ast T$ of
cosmic expansion against the inward restoring force due to the extrinsic
curvature $G$ of our embedded Friedmann $3$-brane $\mathbb{S}_{3}\left(
a_{\#}\right) \subset \mathbb{R}_{4}$.

Much as the radius of a soap bubble reflects a balance between internal
pressure and extrinsic curvature, the value $\kappa =\gamma ^{-2}\frac{1}{2}%
a_{\#}^{2}$ of the gravitational constant is a local \textquotedblleft
memory" of the global balance between pressure and curvature that sets the
equilibrium radius of our Friedmann universe.

Gravity is attractive because neighboring dilation currents $d\varphi ^{0}$
(masses) present \emph{opposite} radiotemporal vorticities%
\begin{equation}
K^{0}\equiv \left( K_{R}^{0}+K_{L}^{0}\right) =\frac{1}{2}\left[ \partial
_{0},\partial _{r}\right] \left( \varphi _{L}^{0}+\varphi _{R}^{0}\right)
e^{0}\wedge e^{r}  \label{50}
\end{equation}%
to each others' worldtube boundaries. To minimize the net vortex energy,
they \emph{advect}, like counter-rotating hydrodynamic vortices \cite{neu};
i.e. attract, orbit around each other---and perhaps fuse.

When massive particles (e.g. protons and neutrons) which contain both
leptonic (light) and baryonic (heavy) spinors get very close together ($%
r\longrightarrow 0$), the interaction term $TrG_{L}\wedge G_{R}\sim r^{-8}$
in the \emph{strong} fields in (\ref{37}) begins to dominate the action (\ref%
{36}). Just as electromagnetic and weak curvatures unify to make up the 
\emph{antiHermitian} ($PT$-antisymmetric, or charge-separated) parts $K_{A}$
of the net spin curvature $2$ form (\ref{37}), \emph{gravitational} and 
\emph{strong} curvatures unify to make the \emph{Hermitian} ($PT$-symmetric,
or neutral) part $K_{H}\oplus G$. We call $\left( K_{H}\oplus G\right) $ the 
\emph{gravitostrong} field.

\section{Gravitomagnetic-Nuclear Couplings and Axial Jets}

The field ($\widehat{\mathbb{M}}$) integral in (\ref{36}) is the action of
the \emph{perturbed vacuum spinors} outside the worldtubes of massive
particles. It gives the standard effective kinetic energy terms for
electroweak and gravitostrong fields \cite{spingeo}. Standard coupling
constants like $\kappa ^{\prime }$ (\ref{49}) are thus determined by the
vacuum-field amplitudes, and hence their values depend on the radius of the
Friedmann solution.

We show below how the field term in action (\ref{36}) contains not only the
standard \emph{Palatini action} for the gravitational field, but predicts 
\emph{a new interaction between gravitomagnetic fields and weak potentials}.
This belongs neither to the electroweak nor gravitostrong sectors, but lives
in the overlap domain demanded by their unification.

The new interaction terms arise by expanding the $\Omega K\Omega $ term in (%
\ref{36}), using identities (\ref{21}). Defining $K\equiv K_{L}+K_{R}\equiv
K_{\alpha \beta }e^{\alpha }\wedge e^{\beta }$ as the \emph{Hermitian} ($PT$%
-symmetric) part of the $gl\left( 2,\mathbb{C}\right) $-valued spin
curvature $2$ form, we obtain%
\begin{equation}
\begin{array}{c}
\frac{i}{2}Tr\left[ \Omega _{L}\wedge K\wedge \Omega _{R}+P\right] \\ 
=\frac{i}{2}Tr\left[ \frac{1}{4}a_{\#}^{-2}q_{\alpha }K^{\alpha \beta }%
\overline{q}_{\beta }+\frac{1}{2}a_{\#}^{-1}q_{\alpha }K^{\alpha \beta }%
\overline{W}_{\beta }+W_{\alpha }K^{\alpha \beta }\overline{W}_{\beta }+P%
\right] \mathbf{d}^{4}v%
\end{array}
\label{51}
\end{equation}%
in $\mathbb{M}_{\#}$ coordinates: The components $K^{\alpha \beta }$ of the 
\emph{dual} spin curvature,%
\begin{equation}
\ast K=\frac{1}{2}K_{\alpha \beta }\epsilon ^{\alpha \beta }\gamma \delta
\;e^{\gamma }\wedge e^{\delta }\text{,}  \label{52}
\end{equation}%
appear because it takes all four $1$ forms and all four Clifford tetrads to
make the invariant volume element, a Clifford \emph{scalar:}%
\begin{equation*}
\sigma _{0}e^{0}\wedge \sigma _{1}e^{1}\wedge \sigma _{2}e^{2}\wedge \sigma
_{3}e^{3}=i\sigma _{0}\mathbf{d}^{4}v=i\left\vert g\right\vert ^{\frac{1}{2}%
}\sigma _{0}\mathbf{d}^{4}V\text{.}
\end{equation*}%
Clifford multiplication is thus \emph{dual} to exterior multiplication via
the scalar product defined by the integral of the wedge product of
Clifford-algebra-valued forms \cite{tqac}.

The term in $a_{\#}^{-2}\left[ qK\overline{q}+P\right] \left\vert
g\right\vert ^{\frac{1}{2}}\mathbf{d}^{4}V$ gives the \emph{Palatini action}
for gravitation \cite{sachs}. Its variation with respect to the tetrads
gives the Einstein field equations \cite{sachs}, with a gravitational
constant of $\kappa ^{\prime }=\gamma ^{-2}\frac{1}{2}a_{\#}^{2}$. The fact
that we get the \emph{same} value as (\ref{49}) above provides an
independent check on our boundary-integral method.

The terms in $a_{\#}^{-1}\left[ qK\overline{W}+P\right] $ predict a new
interaction between Lens-Thirring gravitational fields and weak potentials 
\cite{axialjets}. This is not a \textquotedblleft fifth force", but an
overlap between the standard interactions that lives in the overarching
domain of their unification \cite{spingeo}. These new cross terms predict
new physical phenomena---\emph{axial jets} \cite{axialjets}.

Suppose some weakly-decaying particles are located near the $3$-axis of a
disk, $D$, rotating about the $3$ axis with angular velocity $\omega $. The
graviatational field%
\begin{equation}
\begin{array}{c}
K=K_{03}e^{0}\wedge e^{3}+K_{12}\left( \omega \right) e^{1}\wedge e^{2}\text{%
;} \\ 
\ast K=K_{03}e^{1}\wedge e^{2}-K_{12}\left( \omega \right) e^{0}\wedge e^{3}%
\end{array}
\label{53}
\end{equation}%
then contains the \emph{gravitomagnetic} (Lens-Thirring) component $%
K_{12}^{3}\left( \omega \right) \sigma _{3}=K_{3}^{03}\sigma _{3}$, whose
magnitude depends upon $\omega $. The new $qKW$ cross term%
\begin{equation}
\begin{array}{c}
\frac{i}{2}Tr\Omega _{0}^{\,0}\sigma _{0}e^{0}\wedge K_{12}^{3}\left( \omega
\right) \sigma _{3}e^{1}\wedge e^{2}\wedge \frac{i}{2}\partial _{3}\left[
\zeta _{L}^{3}-\zeta _{R}^{3}\right] \sigma _{3}e^{3} \\ 
=\frac{1}{4}\Omega _{0}^{\,}K^{03}\left( \omega \right) \partial _{3}\left[
\zeta _{L}^{3}-\zeta _{R}^{3}\right] \sigma _{0}d^{4}v\equiv -V\left( 
\mathbf{\omega }\cdot \mathbf{W}\right) \sigma _{0}d^{4}v%
\end{array}
\label{54}
\end{equation}%
gives an effective potential in the dot product of the angular momentum
vector $\mathbf{\omega }e_{3}$ and the nonAbelian vector potential $\mathbf{W%
}$, multiplying the \emph{timelike part of the vacuum effective spin
connection}. On a $3$-brane $S_{3}\left( t\right) $, with (local) expansion
rate \cite{spingeo}, \cite{im}, \cite{xueg}%
\begin{equation}
\frac{\overset{\cdot }{a}}{a}=\frac{\overset{\cdot }{y}^{0}}{a_{\#}}\text{,}
\label{55}
\end{equation}%
this reads\footnote{%
Two pieces of evidence converge on a current (local) value of $\overset{%
\cdot }{y}^{0}=0.16$ (where $c=1$). This predicts values of $\theta
_{W}=28.5^{\circ }$ \emph{and} of $\alpha =\frac{1}{137.6}$ for the Weinberg
angle \cite{xueg} and the fine structure constant \cite{tqac}, which match
the observed values quite closely.}%
\begin{equation}
\widehat{\Omega }_{0}^{\pm }e^{0}=\frac{1}{2a_{\#}}\left( \pm i-\overset{%
\cdot }{y}^{0}\right) \sigma _{0}e^{0}\text{.}  \label{56}
\end{equation}%
Letting the nonAbelian vector potential have both antiHermitian (imaginary)
and Hermitian (real) parts,%
\begin{equation*}
\begin{array}{c}
W_{L}=\frac{i}{2}\partial _{3}\zeta _{L}^{3}\sigma _{3}e^{3}\text{;} \\ 
W_{R}=\frac{i}{2}\partial _{3}\zeta _{R}^{3}\overline{\sigma }_{3}e^{3}%
\end{array}%
\end{equation*}%
where $\zeta ^{3}\left( x\right) \equiv \theta ^{3}\left( x\right) +i\varphi
^{3}\left( x\right) $, we obtain the real cross term%
\begin{equation}
V\left( \mathbf{\omega \cdot W}\right) =-\frac{1}{2a_{\#}}K^{03}\left(
\omega \right) \left[ \overset{\cdot }{y}^{0}\partial _{3}\left( \varphi
_{L}^{3}-\varphi _{R}^{3}\right) \mp \partial _{3}\left( \theta
_{L}^{3}-\theta _{R}^{3}\right) \right]   \label{57}
\end{equation}%
in $\mathcal{L}_{g}$.

In an expanding universe ($\overset{\cdot }{y}^{0}>0$), $V$ decreases when
the left-chirality parts of the weak-decay products are boosted more along
the $\widehat{\mathbf{\omega }}$ direction than the right, independently of
their charges; the second term describes a charge-dependent spin
polarization. The net effect of the $qKW$ interaction term is to cause \emph{%
weak-decay products to be ejected with left-helicity with respect to the
axis of rotation,} producing \emph{axial jets.}

Astronomical observations often show plasma jets ejected along the axes of
rotating quasars, pulsars, and active galactic nucleii \cite{agn}. I don't
know if their helicity has been measured. Left-helical polarization of such
jets would tend to support the NM model.

Now it could be argued that, if observed, the left-helical polarization
could be explained by standard electroweak theory. But left-helicity enters
there as an \emph{assumption.} The NM model \emph{derives} the left-helicity
preference of weak decays from dynamical symmetry breaking of the vacuum
spin connection $\Omega _{0}$ in the \emph{forward} timelike direction: $%
\overset{\cdot }{y}^{0}>0$ in (\ref{57}). If our expanding Friedmann $3$%
-brane $S_{3}\left( t\right) $ is to remain bulk neutral ($PT$-symmetric),
then when $T$ is broken, $P$ must be broken. The $qKW$ term gives a \emph{%
dynamical} mechanism for $P$-symmetry breaking on the cosmological scale.

On microscopic scales, the same $qKW$ mechanism could drive the
left-helicity weak-decay modes of massive spinning particles (e.g. nucleii).
Even when there is no net spin, interaction with the (predominantly left
helicity) vacuum spinors could drive left-helicity weak decays through the
Newtonian field%
\begin{equation*}
K_{0r}=\left( \ast K\right) ^{\theta \varphi }\equiv K^{\theta \varphi }
\end{equation*}%
of a massive particle \cite{strat}, via terms like $q_{\theta }K^{\theta
\varphi }W_{\varphi }$.

$WK\overline{W}$ terms in (\ref{51}) like%
\begin{equation*}
\begin{array}{c}
TrW_{0}K^{03}\overline{W}_{3}+P=Tr\left( W_{0}\otimes _{A}\overline{W}%
_{3}\right) \left( K^{03}\otimes \mathbf{1}\right) \\ 
=Tr\left( W_{0}\overline{W}_{3}-W_{3}\overline{W_{0}}\right) \gamma
_{03}K^{03}\equiv TrG_{03}K^{03}\text{,} \\ 
\text{where\quad }\gamma _{03}\equiv \left[ 
\begin{array}{ccc}
2 & 0 & 0 \\ 
0 & -2 & 0 \\ 
0 & 0 & 0%
\end{array}%
\right] \equiv \gamma _{3}-\gamma _{8}\text{,}%
\end{array}%
\end{equation*}%
seem to predict new interactions between gravitational and \emph{strong} ($%
su\left( 3\right) $-valued) fields: \emph{strong} (nuclear fusion) jets
emitted along the axis of a supermassive rotating \textquotedblleft
nucleus", like a neutron star. Meanwhile, the Newtonian gravitational field $%
\ast K_{0r}=K^{\theta \varphi }$ and antisymmetric tensor products of $%
su\left( 2\right) $ potentials like $W_{1}+iW_{2}\equiv W_{\theta }^{+}$ and 
$\overline{W}_{1}-i\overline{W}_{2}\equiv W_{\varphi }^{-}$ combine to make
new cross terms like%
\begin{equation*}
\begin{array}{c}
TrW_{\theta }K^{\theta \varphi }\overline{W}_{\varphi }+P=TrG_{\theta
\varphi }K^{\theta \varphi }\text{;} \\ 
G_{\theta \varphi }\equiv \left[ W_{\theta }\overline{W}_{\varphi
}+W_{\varphi }\overline{W}_{\theta }\right] \left[ 
\begin{array}{ccc}
0 & 1 & 0 \\ 
-1 & 0 & 0 \\ 
0 & 0 & 0%
\end{array}%
\right] \text{.}%
\end{array}%
\end{equation*}%
These seem to predict accelerated nuclear reaction rates in a strong
Newtonian field. I don't know if such effects are observed, or, if so, how
standard models explain them.

The advantage of the \emph{Spin}$^{c}$-4 model here is that it remains
nonsingular in collapsed matter---the regime of \emph{grand} \emph{%
unification} (see Appendix)---where standard models break down.

\section{The Singular $Spin^{c}$-4 Complex and Quantum Gravity}

Let's return to our central point:

\begin{enumerate}
\item[\thinspace ] \emph{Nonlinear interaction of localized matter spinors
through globallly-nontrivial vacuum spinor fields creates the inertial
masses of the particles. The resulting perturbation to the vacuum spinors
produces the gravitational interaction between particles.}
\end{enumerate}

Let's see how this is implemented on the microscopic scale. Here, the mass
scatterings that define the fuzzy boundary $\partial B_{4}$ of the
particle's world tube are displaced by nonunitary perturbations of the
vacuum spinors sourced in other particles. This is \emph{quantum gravity.}

Quantum field theory is statistical mechanics in imaginary time \cite{grosse}%
, $T\equiv y^{0}$. $T$ \emph{is} cosmic time, which enters kinematically as
the \emph{imaginary} part of a complex time parameter $z^{0}\equiv
x^{0}+iy^{0}$, whose real part $dx^{0}\equiv \left\vert d\mathbf{x}%
\right\vert $ is arclength increment along a null ray (e.g. the path of a
photon). The \emph{energy} of a state (multiplied by the temperature) is
replaced by the \emph{action} of a path (divided by $i$%
h{\hskip-.2em}\llap{\protect\rule[1.1ex]{.325em}{.1ex}}{\hskip.2em}%
) in all ensemble averages.

Note \cite{spingeo} that the 4 spinor fields $\left( \psi _{L-},\psi
_{R-},\psi ^{L+},\psi ^{R+}\right) $ are \emph{analytic:} they obey the
Cauchy-Riemann (CR) equations (\ref{10}). The remaining 4 spinor fields $%
\left( \psi _{L+},\psi _{R+},\psi ^{L-},\psi ^{R-}\right) $ are \emph{%
conjugate analytic}. \emph{It is these analyticity conditions that justify} 
\emph{Wick rotation,} which translates the statistical mechanics of null
zig-zags in \emph{Euclidean spacetime,} with coordinates $\left( y^{0},%
\mathbf{x}\right) $, to Feynman integrals over compactified Minkowsky space $%
\mathbb{M}_{\#}$, with coordinates $\left( x^{0},\mathbf{x}\right) $ \cite%
{grosse}.

A stochastic version of our model, in which the classical action is replaced
by a statistical propagator (the sum over null zig-zags \cite{ord}, \cite%
{penrose}) is thus a theory of quantum gravity, \emph{provided} that the 
\emph{vacuum spinors} that do the mass scatterings that confine the null
zig-zags of massive particles to timelike worldtubes \cite{penrose}, \cite%
{xueg}, are also modelled statistically.

The chiral spinor fields that bind to form a massive particle are \emph{%
lightlike:} each has \emph{definite helicity} \cite{penrose}. Their phases $%
\zeta ^{\alpha }\left( z\right) $ or $\zeta ^{\alpha }\left( \bar{z}\right) $
may propagate \emph{only} along segments of \emph{forward} characteristics, $%
\gamma _{+}:dy^{0}=\left\vert d\mathbf{x}\right\vert $, or backward
characteristics, $\gamma _{-}:dy^{0}=-\left\vert d\mathbf{x}\right\vert $%
\emph{.} To make a massive particle with \emph{definite spin,} $L$- and $R$%
-chirality moieties must be \emph{counterpropagating,} i.e. have opposite
momenta, but the same spins, and therefore \emph{opposite helicities}. The
propagator for a massive bispinor particle \cite{ord}, \cite{penrose} is
thus a \emph{sum over null zig-zags:\ }counterpropagating lightlike segments
with \emph{mass scatterings} at each corner.

These mass scatterings are vertices in a Riemann sum for the action, $S_{g}$%
, which we compute as follows:

\begin{enumerate}
\item[i)] Re-express $S_{g}$ with respect to complex Clifford (internal) and
spacetime (external) \emph{null} \emph{tetrads} on $\mathbb{CM}_{\#}\subset
T^{\ast }\mathbb{M}_{\#}$, 
\begin{equation}
\begin{array}{c}
\sigma _{\pm }\equiv \frac{1}{\sqrt{2}}\left( \sigma _{1}\pm i\sigma
_{2}\right) \text{;\qquad }\sigma _{\uparrow \downarrow }\equiv \frac{1}{%
\sqrt{2}}\left( \sigma _{0}\pm \sigma _{3}\right) \text{,} \\ 
e^{\pm }\equiv \frac{1}{\sqrt{2}}\left( e^{1}\pm ie^{2}\right) \text{;\qquad 
}e^{\uparrow \downarrow }\equiv \frac{1}{\sqrt{2}}\left( e^{0}\pm
e^{3}\right) \text{.}%
\end{array}
\label{58}
\end{equation}

\item[ii)] Discretize the $\mathbf{8}$-spinor form (\ref{27}) for our
Lagrangian density $\mathcal{L}_{g}$, and compute our Riemann sums for $%
S_{g} $ over \emph{null lattice,} $N$, stepped off by null tetrads (\ref{58}%
). The line segments of $N$ are the (lightlike) rays of spinor fields. Its
points (vertices) are \emph{scattering events,} where $m$ incoming chiral
pairs of spinors---some of which may be vacuum spinors---scatter into $%
\left( 4-m\right) $ outgoing pairs. These \emph{Spin}$^{c}$\emph{-4
scatterings} are the discrete versions of \emph{Spin}$^{c}$-4 resonances
like (\ref{31}). The spinor differentials $\mathbf{d}\psi _{I}$ in $\mathcal{%
L}_{g}$ (\ref{27}) are replaced by \emph{first differences}%
\begin{equation*}
\psi _{I}\left( p+\bigtriangleup \right) -\psi _{I}\left( p\right) \equiv
\bigtriangleup \psi
\end{equation*}%
between neighboring lattice points in $\sum $. The \textquotedblleft
conjugate gradient" terms (\ref{9}) are approximated by discrete $gl\left( 2,%
\mathbb{C}\right) $ phase differences:%
\begin{equation}
\psi ^{I}\mathbf{d}\psi _{I}=\frac{i}{2}\mathbf{d}\zeta _{I}^{\alpha }\sigma
_{\alpha }\equiv \frac{i}{2}\mathbf{d}\zeta _{I}\sim \frac{i}{2}%
\bigtriangleup \zeta _{I}\equiv \frac{i}{2}\left( \bigtriangleup \theta
^{\alpha }+i\bigtriangleup \varphi ^{\alpha }\right) \sigma _{\alpha }\text{.%
}  \label{59}
\end{equation}%
Both the scalar ($\mathbb{C}u\left( 1\right) $) parts%
\begin{equation*}
\bigtriangleup \zeta ^{0}\equiv \left( \bigtriangleup \theta
^{0}+i\bigtriangleup \varphi ^{0}\right) \equiv \left( \bigtriangleup \text{%
charge}+i\bigtriangleup \text{energy}\right)
\end{equation*}%
and the vector ($\mathbb{C}su\left( 2\right) $) part%
\begin{equation*}
\bigtriangleup \zeta ^{j}\equiv \left( \bigtriangleup \theta
^{j}+i\bigtriangleup \varphi ^{j}\right) \equiv \left( \bigtriangleup \text{%
spin}+i\bigtriangleup \text{momentum}\right)
\end{equation*}%
contribute to the net $gl\left( 2,\mathbb{C}\right) $ phase increment $%
\bigtriangleup \zeta ^{\alpha }\left( p\right) \sigma _{\alpha }$ at vertex $%
p$. However, to obtain an action that is a \emph{real scalar} under all
passive spin ($E_{P}$) transformations, we restrict our sums to the \emph{%
scalar} $\left( \sigma _{0}\right) $ component $\bigtriangleup \mathcal{L}%
_{g}^{0}\sigma _{0}$ of each \emph{Spin}$^{c}$-$4$\emph{\ }scattering
between $J$ chiral pairs of matter spinors and $\left( 4-J\right) $ vacuum
pairs. The charges $\bigtriangleup \theta ^{0}\left( p\right) $, spins $%
\bigtriangleup \theta ^{j}\left( p\right) $, and $3$ momenta $\bigtriangleup
\varphi ^{j}$ add to zero at every \emph{Spin}$^{c}$-$4$\emph{\ }scattering, 
$p$.

\item[iii)] To automatically insure quantization of action \cite{tqac}, we
should take the area of the elementary $2$ cells $\widehat{\gamma }_{2}$ in
a rectangular null lattice to be $\bigtriangleup p\bigtriangleup q=\frac{%
\text{%
h{\hskip-.2em}\llap{\protect\rule[1.1ex]{.325em}{.1ex}}{\hskip.2em}%
}}{2}$. This not only assures that $\bigtriangleup \zeta \left( \gamma
_{2}\right) =2\pi in$ about \emph{any} phase space cycle $\gamma _{2}$, but
also allows all of the spinor wave functions $\psi \left( p\right) $ to be 
\emph{single-valued} at each $p\in N$. The singular homology $H_{\ast
}\left( N\right) $ should be a skeleton for the homology $H_{\ast }\left( 
\widehat{\mathbb{M}}\right) $ of our spacetime manifold minus the singular
loci \cite{tqac}. Then all of the topological charges---$J$ forms $D^{J}\in
H^{J}\left( \widehat{\mathbb{M}}\right) $ quantized over cycles in $\widehat{%
\mathbb{M}}$---will have discrete realizations as products of $J$ net phase
differences quantized over discrete cycles in $N$. We call this complex of
spinor phase shifts quantized over cycles in the null lattice $N$ a \emph{%
singular Spin}$^{c}$\emph{-4} complex, $\sum $.
\end{enumerate}

The action of each $\sum $ is the sum of contributions from every \emph{Spin}%
$^{c}$-$4$ scattering in $\sum $. \emph{Particle propagators} are sums over
all null zig-zag paths in $\sum $ that connect the initial and final state 
\cite{xueg}.

Now the bending of the worldtube $B_{4}\left( P\right) $ of a massive
particle $P$ is described by a changing $4$ momentum $\bigtriangleup \varphi
^{\alpha }\left( P\right) =-\bigtriangleup \varphi ^{\alpha }\left( V\right) 
$ imparted by mass scatterings on $\partial B_{4}\left( P\right) $ with the
vacuum spinors, $V$. Anholonomic \emph{changes} $\bigtriangleup
\bigtriangleup \varphi ^{\alpha }\left( V\right) \sim G$ in the vacuum
spinors due to \emph{sources} are what impart \emph{curvature} $%
\bigtriangleup \bigtriangleup \varphi ^{\alpha }\left( P\right) $ at the
worldtube boundary $\partial B_{4}\left( P\right) $), via the discrete
version of the boundary integral in (\ref{36}).

Mass scatterings at the worldtube boundary account not only for the
curvature of the worldtube $B_{4}\left( P\right) $, but also for the \emph{%
annihilation} of $P$ and \emph{creation} of intermediate particles by
recombination of its spinor components with each other and with the vacuum
spinors.

Each particle is composed of $J$ \emph{chiral pairs} of $L$- and $R$%
-chirality spinors: leptons of 1, mesons of 2, and hadrons of 3 pairs,
respectively \cite{strat}. Their reactions are \textquotedblleft crossover"
exchanges of spinors from the ingoing to outgoing sets of particles---and
with the $\left( 4-J\right) $ remaining pairs of \emph{vacuum} spinors that
make the spacetime tetrads (\ref{17}), metric tensor (\ref{16}), and the
effective vector potentials (\ref{9}) and fields in (\ref{36}). Particle
propagators, like the Dirac propagator \cite{ord}, thus include creation of
intermediate particles and their return to the vacuum \textquotedblleft
sea". In this sense, the NM model is innately quantum mechanical, and does
not need to be \textquotedblleft quantized".

But wait! There are two basically different recipes, representing different
underlying physical processes, for computing the \textquotedblleft sum over
null zig zags" in the Dirac propagator \cite{ord}!

\begin{enumerate}
\item[R1)] \textbf{The \textquotedblleft sum over histories".} Create an
ensemble $\Sigma _{C}$ of singular $Spin^{c}$-4 complexes $\left\langle
\Sigma _{C}\right\rangle $ with the same set of topological charges and
particle trajectories, $C$, as our classical \emph{Spin}$^{c}$\emph{-4
complexes} and with isomorphic singular homologies. Each $\Sigma _{C}$
represents a different microscopic history of mass scatterings and
intermediate annhialation/creation events compatible with the classical
history, $C$. \emph{Now average over the ensemble of all such }$\Sigma _{C}$ 
\emph{to get quantum mechanical expectation values of observables}---just as
we average over all microstates to get expectations values in statistical
mechanics. These expectation values, $\left\langle \psi \right\rangle \left(
p\right) $, evolve according to the Dirac propagator.

\item[R2)] \textbf{The stochastic process.} Create a random walk of the
(lightlike) rays of matter spinors, $\psi $, and their mass-scatterings
vertices, $p$, with the remaining \emph{vacuum spinors, treated as random
fields,} with mean distributions (\ref{18}), and variances $\left\vert
\bigtriangleup \mathbf{\varphi }\right\vert \left\vert \bigtriangleup \theta
\right\vert =\frac{\text{%
h{\hskip-.2em}\llap{\protect\rule[1.1ex]{.325em}{.1ex}}{\hskip.2em}%
}}{2}$. The resulting probability vector $\psi \left( p\right) $ for the
matter fields evolves by the Dirac propagator.
\end{enumerate}

In R1), there are \textquotedblleft many worlds" that exist simultaneously.
In R2), there is only one world---the \emph{real} \emph{Spin}$^{c}$-4
complex $\sum $---but we cannot know enough information about the vacuum
spinors and their fluctuations to distinguish it from the other members $%
\Sigma _{C}$ of its ensemble. Both remain valid interpretations of NM model,
as they are for standard quantum mechanics, and even for statistical
mechanics!

Suffice it to say here that a realistic model, with one real \emph{Spin}$%
^{c} $-4 complex $\sum $ is not precluded, because the NM model is \emph{%
nonlocal.} \emph{Half} of the spinors incident on every vertex $p\in \sum $
in the \emph{Spin}$^{c}$-4 complex---the conjugate-analytic spinors $\psi
\left( \overline{z}\right) $---propagate \emph{backwards} in cosmic time, $T$%
, i.e. in the \emph{opposite} direction to cosmic expansion. These carry
only slightly less energy than the forward-propagating, analytic spinors $%
\psi \left( z\right) $. Macroscopically \cite{davies}, the net result is
that our Friedmann $3$-simplex $\Sigma _{3}\subset \sum $ expands much more
slowly than the speed of light. Microscopically, the result $R$ of a
measurement may propagate \emph{backwards} in $T$ to become one of the phase
increments $\bigtriangleup \zeta _{R}$ incident on $p$. $R$ would thus
\textquotedblleft bootstrap" cycles in $\sum $ connecting $p$ and $R$%
---self-consistent causal cycles that are temporally \emph{nonlocal.}
\textquotedblleft Paradoxical" cycles (like killing your own grandfather)
would simply \emph{not} boot-strap, and therefore not exist in the \emph{real%
} \emph{Spin}$^{c}$-4 complex. Contrapositively, all the \emph{existing}
vertices $p$ and 1, 2, 3, and 4 cycles $\gamma _{J}$ must be embedded
self-consistentlly in the \emph{real} \emph{Spin}$^{c}$-4 complex---our
whole world, how it was made, and what we make of it.

Any attempt to isolate a local simplex of $\sum $, and give a deterministic
recipe for its $T$ evolution, necessarily ignores links with other local
simplices on its boundary, and with the global, but stochastic vacuum spinor
fields. The best we can do is to average over the ensemble $\Sigma _{C}$ of
simplices with every possible configuration of boundary fields and vacuum
spinors---and so our theory can only predict ensemble averages.

\section{Conclusion}

The vacuum spinors are what produce the inertial mass of a bispinor particle
because mass scatterings with the vacuum fields are what confine chiral
pairs of matter spinors to a timelike worldtube, $B_{4}$. Inside $B_{4}$,
the matter spinors \textquotedblleft zag" backwards in cosmic time $T$ \emph{%
almost} as often as they \textquotedblleft zig" forward. Macroscopically,
the timelike flux of the \emph{spinfluid} decreases away from the boundary $%
\partial B_{4}$ of a source. This creates a \textquotedblleft curl" $K_{0r}$
in the vacuum flow that causes the worldtubes of test particles to curve,
i.e to accelerate towards the source.

Spinfluid models thus give a \emph{mechanism} for gravitation. Qualitatively,

\begin{enumerate}
\item[i)] vorticity arises on the boundary of each energy-momentum current, $%
\ast T$.

\item[ii)] rotation $\mathcal{R}$ of the spin fluid flow falls with distance
from the source current. Its moment $\ast G$ at the boundary of the
worldtube of a test current causes this worldtube to bend towards the
source. Thus,

\item[iii)] neighboring centrifugal currents (masses) present \emph{opposite}
radiotemporal vorticities $G_{or}$ to each other's worldtube
boundaries---and therefore \emph{attract} (or advect, like hydrodynamic
vortices \cite{neu}).
\end{enumerate}

Quantitatively, the gravitational constant is determined by the same balance
between outward flow and boundary curvature that determines the equilibrium
radius of the Friedmann universe. The power of spinfluid models to determine
some constants of nature\ also makes them \emph{falsifiable. }For example,
relations (\ref{29}) and (\ref{47}) give the value 
\begin{equation}
\kappa m_{e}^{2}=\frac{\gamma ^{-2}}{8}\left( T\right)  \label{60}
\end{equation}%
for the dimensionless constant that measures the ratio of gravitational to
electromagnetic forces between electrons on our dilated Friedmann $3$-brane, 
$\mathbb{S}_{3}\left( a\right) $. The predicted value of (\ref{60}) would
match the value of $10^{-39}$ observed today with a dilation factor of $%
\gamma \sim 10^{20}$ and predicts that this ratio should \emph{decrease}
with cosmic expansion!

At $T=0$, on $\mathbb{S}_{3}\left( a_{\#}\right) $, gravitational and
electromagnetic forces have the same order of magnitude, because they have 
\emph{not yet separated}. It is \emph{cosmic expansion}---the breaking of $T$%
--reflection symmetry---that divides the forces into electroweak ($PT$%
-antisymmetric) and \emph{gravitostrong} ($PT$-symmetric) sectors \cite%
{spingeo}.

Microscopically, the action $S_{g}$ is a sum over mass scatterings between $%
J $ chiral pairs of matter spinors and $\left( 4-J\right) $ vacuum pairs
plus the action of the perturbed vacuum spinor fields. Together, these make
up a \emph{Spin}$^{c}$-4 complex. The statistical mechanics in
\textquotedblleft imaginary time" $T\equiv y^{0}$ of all null \emph{Spin}$%
^{c}$-4 complexes $\Sigma _{C}$ compatible with a given set $C$ of classical
particle trajectories, masses, and charges is \emph{quantum gravity.}

I don't know how our NM model compares with other theories of quantum
gravity. There is a crucial experimental test, however. The NM model
predicts a new effect in \emph{supervortical} regimes: that Lens-Thirring
fields should polarize weak decays, producing \emph{left-helicity} jets
about the rotation axis \cite{spingeo}, \cite{xueg}. Perhaps such axial jets
could be measured in terrestrial laboratories or in astronomical
observations.

More significant than the prediction of new effects of the derivation of
fundamental constants and selection rules or are the qualitative features of
the class of \emph{quantum} \emph{spinfluid} models that enable them to
reconcile quantum mechanics and general relativity. These are

\begin{enumerate}
\item A Lagrangian density with \emph{no} free parameters that is a \emph{%
natural} $4$ form---i.e. invariant under the group of passive spin
isometries in curved spacetime, including symplectomorphisms.

\item An action that \emph{includes} a bounded vacuum energy which depends
on the radius $a\left( t\right) $ of the Friedmann solution. This breaks
dilation invariance and sets all length and mass scales. It includes a
repulsive term at high densities that prevents collapse to a singularity.

\item Values for the standard coupling constants that are either
\textquotedblleft frozen in\textquotedblright\ by the history of dynamical
symmetry breaking, or that depend on cosmic time $T$ through the radius $%
a\left( T\right) $.

\item Effective electroweak, strong, and gravitational \emph{field} actions,
along with minimal coupling through spin connections in the covariant
derivatives.

\item Fields which are sourced in localized currents with \emph{%
topologically quantized charges}. These charges parallel the electric
charges, masses, spins, and lepton or baryon numbers of the observed
families of particles.

\item A \emph{mechanism} for gravitation derived from the \emph{same}
nonlinear coupling to the vacuum spinors that creates the \emph{inertial
masses} of particles.

\item Quantum propagators, including intermediate creation and annihilation
operators, which are \emph{derived }from the statistical mechanics of null
zig-zags of the (lightlike) spinor fields that weave the (timelike)
worldtubes of massive particles and the (spacelike) fabric that connects
them.
\end{enumerate}

The NM model is the minimal model with these features, because:

\begin{enumerate}
\item[a)] It takes the intersection of $4$ null cones to determine a point
on $\mathbb{M}$.

\item[b)] Each nullcone is generated, via $S^{-1}$ of (\ref{15}), by the
product (\ref{9}) of $1$ cospinor and 1 spinor differential. There must be $%
\mathbf{8}$ spinor fields in all; $\mathbf{4}$ $PTC$-equivalent pairs.

\item[c)] For \emph{symplectic} invariance, each term in $\mathcal{L}_{g}$
must contain $4$ spinors $\Psi ^{I}$ and $4$ differentials $\mathbf{d}\Psi
_{I}$. The symplectic forms are the $4$ Maurer-Cartan $1$ forms $\Psi ^{I}%
\mathbf{d}\Psi _{I}=\Omega _{I}$. Their integrals are quantized over $1$%
-cycles.

\item[d)] Under $PTC$ symmetry, $\mathcal{L}_{g}$ reduces to the
Maurer-Cartan (MC) $4$ form. Its action $S_{g}$ measures the covering number
of the compactified internal group $U\left( 1\right) \times SU\left(
2\right) $ over the compactified spacetime manifold $\mathbb{M}_{\#}$. $%
S_{g} $ is a topological invariant; it comes in integral units.

\item[e)] Since the vacuum energy has a topological upper bound of $16\pi
^{3}W$ \cite{tqac}, with $W$ an integer that does not change with refinement
of the lattice, there are no built-in divergences, and we don't have to
worry about unbounded vacuum energies rolling up our space to a point.\
There is no need for renormalization.

\item[f)] The NM model admits \emph{nonperturbative} solutions \cite{tqac}, 
\cite{im} in the \emph{superdense} regimes inside collapsed objects like
neutron stars, black holes, and the early universe. These solutions are
regularized by the resistance (\ref{26}) of the topologically nontrivial
vacuum to compression to a point \cite{tqac}---a \textquotedblleft pressure"
of the vacuum spinors that contributes no additional energy-momentum (see
Appendix).
\end{enumerate}

The vacuum energy---or \emph{dark energy}---is simply the energy of the
homogeneous distribution of vacuum spinor fields $\left\{ \hat{\psi}_{I},%
\hat{\psi}^{I}\right\} $, on which the gauge fields ride like waves on the
surface of an ocean. Material particles are the \emph{caustics} of these
vacuum spin waves \cite{tqac}, \cite{strat}---topological dislocations with
quantized charges.

We have found some classical soliton solutions \cite{preprint} for the
unified field action $\mathcal{L}_{g}$ with topologically quantized charges 
\cite{tqac}. These exemplify the inner solutions of Section 4. A
stratification lemma \cite{strat} classifies the varieties of inner
solutions that can exist, and their allowed reactions. These seem to
parallel the observed particles and reactions, with \emph{no exceptions}%
\footnote{%
When I presented the NM model at CERN in July of 2000, T.T. Wu asked,
\textquotedblleft Can you account for neutral pion decay?" Wu's question was
the gateway for me into spinor exchanges with the vacuum sea. It then became
clear how even the simplest reactions and interactions, like photon emission
and absorption, recruit spinors from the vacuum and then return them \cite%
{strat}.} noted so far.

The above results all support our picture of the \emph{Spin}$^{c}$-4 complex
of global vacuum spinors, matter spinors, and their mass scatterings \emph{as%
} the microscopic reality. Spacetime, gauge fields, and the observed
varieties of matter fields and their reactions, including gravitation, \emph{%
emerge} from dynamical stratification \cite{strat} of the $Spin^{c}$-4
complex as $1$, 2, 3, and 4 chiral pairs of spinor fields break away from
the $PTC$-symmetric vacuum distribution.

\section{Acknowledgments and Dedication}

Thanks go to Jaime Keller \cite{keller}, who first understood and explained
to me the advantages of the chiral $gl\left( 2,\mathbb{C}\right) $
presentation over the twistor presentation of the conformal group for
massive bispinor particles. I am very grateful to T.T. Wu for listening
carefully to my presentation of the NM model at CERN in July, 2000---and for
setting me the riddle of neutral pion decay. Thanks go to Astri Kleppe for
first helping me to draw the correspondence of chiral bispinor pairs and
exchanges among the $\mathbf{8}$ spinors to real particles and reactions.
Eternal thanks go to Elaine and Mikaela Cohen for keeping my feet grounded
on this Earth.

This paper is dedicated to the memory of my mother, Florence Channock Cohen,
whose spirit returned from timelike weave of matter into eternal light on
January 19, 2003.

\section{Appendix: Conformal Symmetry Breaking and Particle Nucleation}

Time evolution preserves topological invariants. Like kinks and knots in a
hose as it is pulled tight, topological defects in the spinfluid localize
with cosmic expansion, and they acquire \emph{mass}.

Like the residues of complex scalar fields, our action integral has a nested
set of complex \emph{quaternionic} (Clifford) \emph{residues} \cite{tqac}:
integrals over codimension-$J$\emph{\ singular loci }$D^{J}\subset \mathbb{C}%
_{4}\subset T^{\ast }\mathbb{M}$ in phase space, where the $PTC$-symmetric
geometrical optics ansatz (\ref{1}) breaks down. On cycles $\gamma
_{4-J}\equiv \ast D^{J}$, $J=$ 1, 2, 3, or 4 chiral pairs $\widetilde{\Omega 
}$ \textquotedblleft de-loop" from their global vacuum distributions (\ref%
{18}) and appear as localized \emph{matter spinors}, quantized over closed ($%
4-J$)-branes. They thus acquire \emph{masses} and topological charges. The
action of the $W$ chiral pairs of \emph{vacuum spinors} topologically
trapped on $\widehat{\mathbb{M}}$ integrates to a constant, $-16\pi ^{3}W$,
because their intensities $\left\vert \Psi _{I}\right\vert ^{2}$ scale as $%
\gamma ^{-4}$, and the volume element scales as $\gamma ^{4}$. However, the
action of each \emph{matter-spinor} pair topologically trapped on a
codimension-$J$ cycle $D^{J}$ scales as $\gamma ^{J}$ when integrated over
the $J$ orthogonal directions. Since the matter spinors in supports $\gamma
_{4-J}\equiv \ast D^{J}$ meld continuously into the surrounding vacuum $%
\widehat{\mathbb{M}}$, their supports $\gamma _{1}$, $\gamma _{2}$, $\gamma
_{3}$, and $\gamma _{4}$ become $J=$1, 2, 3, and 4-cycles---over which their
phase differentials (Lie-algebra-valued forms) $\widetilde{\Omega }^{1}$, $%
\widetilde{\Omega }^{2}$, $\widetilde{\Omega }^{3}$, and $\widetilde{\Omega }%
^{4}$ are \emph{quantized }\cite{tqac}---and so their integrals do not
change with $\gamma $:%
\begin{equation}
\begin{array}{c}
\int_{\widehat{\mathbb{M}}}\widehat{\Omega }^{4}=16\pi ^{3}W\text{;} \\ 
\underset{D^{1}}{\sum }\int_{S_{3}\times \gamma _{1}}\widehat{\Omega }%
^{3}\wedge \widetilde{\Omega }^{1}=i8\pi ^{2}\gamma \underset{D^{1}}{\sum }%
i2\pi m\equiv -16\pi ^{3}M\gamma \\ 
\underset{D^{2}}{\sum }\int_{S_{2}\times \gamma _{2}}\widehat{\Omega }%
^{2}\wedge \widetilde{\Omega }^{2}=4\pi \gamma ^{2}\underset{D^{2}}{\sum }%
4\pi q\equiv 16\pi ^{3}Q\gamma ^{2} \\ 
\underset{D^{3}}{\sum }\int_{S_{1}\times \gamma _{3}}\widehat{\Omega }%
^{1}\wedge \widetilde{\Omega }^{3}=i2\pi \gamma ^{3}\underset{D^{3}}{\sum }%
i8\pi ^{2}B\equiv -16\pi ^{3}N\gamma ^{3}\text{.}%
\end{array}
\label{61}
\end{equation}%
Note that the topological charges of the matter spinors become quantized
over the \emph{dual} cycles $\gamma _{4-J}$ spanned by the $4-J$ pairs of 
\emph{vacuum spinors,} $\widehat{\Omega }^{4-J}$ \cite{tqac}.

Elsewhere \cite{tqac}, \cite{strat} we have identified%
\begin{equation*}
\begin{array}{c}
M\text{ as the number of \emph{leptons} (weighted by charge),} \\ 
Q\text{ as the total \emph{charge,} and} \\ 
N\text{ as the number of \emph{hadrons} (weighted by charge)}%
\end{array}%
\end{equation*}%
contained in $\mathbb{M}$.

The net effect of (\ref{61}) is to produce a self-potential well $V\left(
\gamma \right) $ in which the scale factor $\gamma \left( t\right) $ moves
in Minkowsky time, $t$.

Both the vacuum and matter spinors acquire some \emph{kinetic energy}
relative to our expanding $3$-brane $\mathbb{S}_{3}\left( T\right) $, as it
sweeps through their rest frames. Thus the Friedmann action $S\left( \gamma ,%
\overset{\cdot }{\gamma }\right) \equiv T\left( \overset{\cdot }{\gamma }%
\right) -V\left( \gamma \right) $ contains a kinetic energy term $T\left( 
\overset{\cdot }{\gamma }\right) $ in the expansion rate $\overset{\cdot }{%
\gamma }$. To obtain $T\left( \overset{\cdot }{\gamma }\right) $, we
re-express our action integral in a Lorenz frame comoving with the Friedmann
flow:%
\begin{equation}
\begin{array}{c}
\left( E^{0},E^{1},E^{2},E^{3}\right) =\gamma ^{4}\beta \left( \mathbf{d}%
y^{0},e^{1},e^{2},e^{3}\right) \text{;} \\ 
\beta =\left( 1+a_{\#}^{2}\overset{\cdot }{\gamma }^{2}\right) ^{\frac{1}{2}%
}\Longrightarrow \mathbf{d}^{4}V^{\prime }=\left( 1+a_{\#}^{2}\overset{\cdot 
}{\gamma }^{2}\right) ^{2}\gamma ^{4}\mathbf{d}^{4}v\equiv \gamma ^{\prime 4}%
\mathbf{d}^{4}v\text{,}%
\end{array}
\label{62}
\end{equation}%
where $a_{\#}\overset{\cdot }{\gamma }\equiv \overset{\cdot }{y}^{0}$ is the
expansion rate of the logradius (cosmic time) $T\equiv y^{0}$ with Minkowsky
time (arctime) $t\equiv x^{0}$.

We then integrate over $\mathbb{M}^{\prime }\equiv \left( T,\mathbb{S}%
_{3}\left( T\right) \right) \subset \mathbb{C}_{4}$, our expanding
hyperspherical shell thickened up by cosmic (\textquotedblleft imaginary")
time $T$ to obtain the action $S\left( \gamma ,\overset{\cdot }{\gamma }%
\right) $ it contains. We only need to keep track of the \emph{scalar} ($%
\sigma _{0}$) terms in $\mathcal{L}_{g}$---because only Clifford \emph{%
scalars} contribute to the \emph{Trace.}

Next, we expand the scalar $\left( \sigma _{0}\right) $ terms in $S$ by
orders in the net conformal weight, $\gamma ^{p}$. Altogether, we get terms
with conformal weights $p=\left[ -4,-3,-1,0\right] $ in our effective
Lagrangian $\widehat{\mathcal{L}}_{g}\left( \psi ,\mathbf{d}\psi \right) $
for our Friedmann vacuum $\widehat{\mathbb{M}}$ with its embedded
topological twists: spacetime, with a \textquotedblleft foam" of singular
loci---the massive particles. These integrate to terms with conformal
weights, $\gamma ^{p}$: $p=\left[ 0,1,2,3,4\right] $. As $\gamma $ grows,
heavier particles begin to dominate the action.

But, what cuts out our expanding space $\mathbb{M}^{\prime }=\left(
T,S_{3}\left( T\right) \right) \subset \mathbb{C}_{4}$ in the ambient phase
space is the \emph{bulk neutrality condition,} $Q=O$ in (\ref{61}). This
leaves us with terms in $\gamma $ and $\gamma ^{3}$ in our potential energy $%
V\left( \gamma \right) $. Meanwhile, the $a_{\#}\overset{\cdot }{\gamma }=%
\overset{\cdot }{y}^{0}$ term in the volume expansion ratio in the \emph{%
Lorenz} frame comoving with the Friedmann flow provides the kinetic energy
term $T\left( \overset{\cdot }{\gamma }\right) $, giving%
\begin{equation}
S\left( \gamma ,\overset{\cdot }{\gamma }\right) =16\pi ^{3}\left(
1+a_{\#}^{2}\overset{\cdot }{\gamma }^{2}\right) ^{\frac{1}{2}}\left[
W-M\gamma -N\gamma ^{3}\right]  \label{63}
\end{equation}%
for the action contained in $\mathbb{M}^{\prime }$. Defining%
\begin{equation*}
\tau \equiv \left( 1+a_{\#}^{2}\overset{\cdot }{\gamma }^{2}\right) ^{\frac{1%
}{2}}t
\end{equation*}%
as the comoving time increment, then varying $S\left( \gamma ,\overset{\cdot 
}{\gamma }^{\prime }\right) $ with respect to $\gamma $ and $\gamma ^{\prime
}\equiv \partial _{\tau }\gamma $ gives%
\begin{equation}
a_{\#}^{2}\gamma ^{\prime \prime }=\partial _{\gamma }\ln \left[ W-M\gamma
-N\gamma ^{3}\right] \equiv -\partial \gamma V\left( \gamma \right)
\label{64}
\end{equation}%
as the differential equation for the scale factor, $\gamma $.

The radius $a\left( \tau \right) \equiv \gamma \left( \tau \right) a_{\#}$
of our Friedmann 3-brane moves in a self-potential well%
\begin{equation*}
V\left( \gamma \right) \equiv -\ln \left[ W-M\gamma -N\gamma ^{3}\right]
\end{equation*}%
with an equilibrium value of%
\begin{equation}
\gamma _{\#}=\left( \frac{-M}{3N}\right) ^{\frac{1}{2}}\text{.}  \label{65}
\end{equation}%
$\gamma _{\#}$ is \emph{unstable} (inflationary) for $M>0$, $N<0$, and \emph{%
stable} (oscillatory) for $M<0$, $N>0$. \textquotedblleft Natural selection"
seems to prefer a universe with negatively-charged leptons and positive
hadrons!

In the NM model \cite{spingeo}, \cite{tqac}, \cite{strat} a \emph{hadron} is
a bound configuration of $3$ out of the $4$ chiral pairs of spinor fields; a
lepton is one bound bispinor pair. Since a homotopy may split one
codimension-3 singular locus $D^{3}$ into a tensor product of three
codimension-1 locii $D^{1}$, the ratio $M=3N$ is \emph{generic} (homotopy
invariant). This gives $\gamma _{\#}=1$ in (\ref{65}). The compactification
radius $a_{\#}$, which entered \emph{kinematically} \cite{qs} as the
quantization length (\ref{4}) for global $u\left( 1\right) \times su\left(
2\right) $ phase gradients $d\theta ^{\alpha }\sigma _{\alpha }$ in (\ref{4}%
), turns out to be the \emph{dynamical} equilibrium radius of our Friedmann
3-brane.

Interestingly enough, $2a_{\#}$ also turns out to be the \emph{Compton
wavelength of the electron,} $2a_{\#}=m_{e}^{-1}$, as we show in Section 4
of the text.

Note that to prevent the argument for the $\ln $ in $V\left( \gamma \right) $
from going negative, $W$ must exceed%
\begin{equation}
W_{\#}\equiv \frac{-2M}{3}\left( \frac{-M}{3N}\right) ^{\frac{1}{2}}\text{.}
\label{66}
\end{equation}%
$W_{\#}$ is the lower bound on the number of chiral pairs of vacuum spinors
needed to prevent a singularity in the time evolution of the radius $a\left(
\tau \right) $ of the Friedmann solution. The positive sign on $W$ indicates 
\emph{left helicity} of the vacuum spinors \cite{qs}; i.e. $su\left(
2\right) _{L}$ twist along null rays. Thus we detect left-helicity bispinors
like $v_{e}\equiv \left( \xi _{+}\oplus \eta _{-}\right) $; not $\overline{v}%
_{\ell }\equiv \left( \xi _{-}\oplus \eta _{+}\right) $, in the vacuum
spinor flux.

In a universe composed of \emph{matter} ($M<0$, $N>0$) rather than
antimatter, a stable equilibrium $\gamma _{\#}$ is produced when the
aggregative (gravitostrong) force between massive hadrons ($N$) is balanced
by the quantum-mechanical preference of leptons ($M$) for \emph{%
delocalization. }This balance lives in the domain of
electroweak-gravitostrong unification with a \emph{regular,}
topologically-nontrivial distribution of \emph{vacuum spinors}; it \emph{%
cannot} be captured by ad-hoc cutting and pasting of quantum field theory
and general relativity.

\end{document}